# Investigating the inner structure of focal adhesions with single-molecule localization microscopy

H. Deschout[1], I. Platzman[2], D. Sage[3], L. Feletti[1], J. P. Spatz[2] and A. Radenovic[1]

[1]*Laboratory of Nanoscale Biology, Institute of Bioengineering, School of Engineering, EPFL, Lausanne, Switzerland*

[2]*Max-Planck-Institute for Medical Research, Dept. of Cellular Biophysics, University of Heidelberg, Dept. of Biophysical Chemistry, Heidelberg, Germany*

[3]*Biomedical Imaging Group, School of Engineering, EPFL, Lausanne, Switzerland*

**Abstract**

**Cells rely on focal adhesions (FAs) to carry out a variety of important tasks, including motion, environmental sensing, and adhesion to the extracellular matrix. Although attaining a fundamental characterization of FAs is a compelling goal, their extensive complexity and small size, which can be below the diffraction limit, have hindered a full understanding. In this study we have used single-molecule localization microscopy (SMLM) to investigate integrin β3 and paxillin in rat embryonic fibroblasts growing on two different extracellular matrix-representing substrates (i.e. fibronectin-coated substrates and specifically bio-functionalized nano-patterned substrates). To quantify the substructure of FAs, we developed a method based on expectation maximization of a Gaussian mixture that accounts for localization uncertainty and background. Analysis of our SMLM data indicates that the structures within FAs, characterized as a Gaussian mixture, typically have areas between 0.01 and 1 $\mu m^2$, contain 10 to 100 localizations, and can exhibit substantial eccentricity. Our approach based on SMLM opens new avenues for studying structural and functional biology of molecular assemblies that display substantial varieties in size, shape, and density.**

# 1. Introduction

Focal adhesions (FAs) are subcellular macromolecular assemblies consisting of dynamic protein complexes that are localized near the cell membrane. FAs affect nearly all aspects of a cell's life, including, but not limited to, adhesion, directional migration, cell proliferation, differentiation, survival, and gene expression (1). Despite having been studied for several decades, the inner architecture of FAs is still not completely understood. In part, this is due to the limitations of conventional fluorescence microscopy for FA analysis. FAs are molecularly diverse structures, containing a large number of proteins (2). Therefore, their investigation requires imaging techniques that offer sufficient multiplexing capabilities (3). Moreover, FAs have a size that is typically in the order of microns or less, and therefore their internal spatio-temporal organization is not fully resolvable with conventional microscopy.

During the last decade, several super-resolution microscopy techniques have been employed to image FAs (4-9). An important insight from these studies was that FAs are not homogeneous spatial structures. Initially, photo-activated localization microscopy (PALM) was used to reveal that FAs can consist of patches of proteins with submicron dimensions (4, 9). Later on, Bayesian localization microscopy and structured illumination microscopy showed that many FAs exhibit discontinuous elongated (or fiber-like) substructures (5, 6). Moreover, single-particle tracking demonstrated that proteins can diffuse within FAs (7, 8), which again suggests that they have an internal spatial organization. However, these have all been qualitative observations, and a quantitative analysis of the FA substructure is still lacking.

For quantitative analysis of the internal spatial organization of FAs, single-molecule localization microscopy (SMLM) can potentially be implemented (10). SMLM data consists of the localizations of individual photo-activatable or photo-switchable fluorescent molecules. Therefore, a variety of methods have been developed to identify and characterize clusters of such localizations (11, 12). These methods are often applied to investigate clusters of receptors in the cell membrane. Such clusters are usually radially symmetric, spatially well separated, and homogeneous in size and density. FA substructures, on the other hand, cannot be characterized as simply. Indeed, adhesions structures can vary from sub-diffraction entities composed of a couple of different proteins (e.g. focal complexes or nascent adhesions) to assemblies of many proteins measuring several microns (e.g. FAs) (13). Moreover, FA subunits are densely packed, since they cannot be resolved using a conventional microscope. Finally, FAs usually have an elongated shape, and the same is possibly true for their subcomponents. Therefore, it is not clear whether established SMLM clustering methods are suitable for the identification of FA substructures.

In this study we have designed a novel approach to investigate the FA substructure. We used expectation maximization of a Gaussian mixture (EMGM) (14) to interpret SMLM data in terms of spatial probability distributions. EMGM allows to quantify the properties of closely packed localization patterns that exhibit substantial varieties in size, density, and shape, and is therefore well suited for studying the inner architecture of FAs. Importantly, we improved the classical EMGM framework to account for localization uncertainties and the presence of a localization background, both being ubiquitous in SMLM data.

The other goal of this study was to quantify the properties of the subunits of which FAs are composed. For this purpose we used PALM, an implementation of SMLM that is popular for imaging FAs (4, 9, 15-17), since it makes use of photo-activatable fluorescent proteins that can be genetically expressed. More in particular, we used PALM to image integrin β3 and paxillin in fixed rat embryonic fibroblasts (REFs), a well-known cell line for FA investigation. Cell experiments were performed using fibronectin-coated substrates and specifically bio-functionalized nano-patterned substrates, on which ordered patterns of nanoscale adhesive spots were provided (18, 19). Such nano-patterned substrates have already been used to indirectly probe the behavior of FAs on the nanoscale (20). In this way, the spatial organization of integrin binding sites is precisely controlled, ensuring that the observed substructures are innate to FAs. Application of our improved version of EMGM on the PALM data allowed us to determine that FAs are composed of structures with areas between 0.01 and 1 $\mu m^2$, containing 10 to 100 localizations, and exhibiting substantial eccentricities down to 0.1.

## 2. Materials and Methods

### 2.1 Microscope

PALM imaging was carried out on a custom built microscope (21, 22). A 50 mW 405 nm laser (Cube, Coherent), a 100 mW 488 nm laser (Sapphire, Coherent), and a 100 mW 561 nm laser (Excelsior, Spectra Physics) were used for excitation/activation. The three lasers were focused into the back focal plane of the objective mounted on an inverted optical microscope (IX71, Olympus). We used a 100× objective (UApo N 100×, Olympus) with a numerical aperture of 1.49 configured for TIRF. A dichroic mirror (493/574 nm BrightLine, Semrock) and an emission filter (405/488/568 nm StopLine, Semrock) were used to separate fluorescence and illumination light. The fluorescence light was detected by an EMCCD camera (iXon DU-897, Andor). An adaptive optics system (Micao 3D-SR, Imagine Optics) and an optical system (DV2, Photometrics) equipped with a dichroic mirror (T565lpxr, Chroma) were placed in front of the EMCCD camera.

### 2.2 Imaging procedure

Cells were imaged in PBS at room temperature. Prior to imaging, 100 nm gold fiducial markers (C-AU-0.100, Corpuscular) were added to the sample for lateral drift monitoring. Axial drift correction was ensured by a nanometer positioning stage (Nano-Drive, Mad City Labs) driven by an optical feedback system (21). Excitation of the mEos2 done at 488 or 561 nm with ~10 mW power (as measured in the back focal plane of the objective). The mEos2 was activated at 405 nm with ~2 mW power. The gain of the EMCCD camera was set at 100 and the exposure time to 50 ms. For each experiment 10,000 camera frames were recorded.

### 2.3 Substrate preparation

Quasi-hexagonal patterns of AuNPs were prepared on 25 mm diameter microscope cover slips (#1.5 Micro Coverglass, Electron Microscopy Sciences) by means of BCML as previously described (18, 19, 23) (Supporting Material). Fibronectin-coated cover slips were prepared by first cleaning with an oxygen plasma and then incubating with PBS containing 50 µg/ml fibronectin (Bovine Plasma Fibronectin, Invitrogen) for 30 minutes at 37 °C. To remove the excess of fibronectin, the cover slip was washed with PBS before seeding the cells.

### 2.4 Cell culture and fixation

The REFs (CRL-1213, ATCC) were grown in DMEM supplemented with 10% fetal bovine serum, 1% penicillin-streptomycin, 1% non-essential amino acids and 1% glutamine, at 37 °C with 5% $CO_2$. The cells were transfected by electroporation (Neon Transfection System, Invitrogen), which was performed on ~$10^6$ cells using 1 pulse of 1350 V lasting for 35 ms. The amount of DNA used for the transfection was 4 µg for both the mEos2-paxillin-22 vector and the mEos2-Integrin-β3-N-18 vector. Around $2.10^5$ transfected cells were seeded on individual cover slips and grown in cell culture medium without penicillin-streptomycin, at 37 °C with 5% $CO_2$. The cells were washed with PBS around 20h after transfection (Fig. S1 in the Supporting Material), and then incubated in PBS with 2.5% paraformaldehyde at 37 °C for 10 minutes. After removing the fixative, the cells were again washed with PBS, and the cover slip was placed into a custom made holder.

## 2.5 PALM data analysis

The recorded images were analyzed by a custom written algorithm (Matlab, The Mathworks) that was adapted from a previously published algorithm (4, 22). First, peaks were identified in each camera frame by filtering and applying an intensity threshold. Only peaks with an intensity at least 4 times the background were considered to be emitters. Subsequently, the peaks were fitted by maximum likelihood estimation of a 2D Gaussian distribution (24). Drift was corrected in each frame by subtracting the average position of the fiducial markers from the positions of the emitters in that frame. The localization uncertainty for each emitter was obtained from the Cramér-Rao lower bound of the maximum likelihood procedure (25). PALM images were generated by plotting a 2D Gaussian centered on each fitted position with a standard deviation equal to the corresponding localization uncertainty. Only positions with a localization uncertainty between 0 and 40 nm were used.

## 2.6 EMGM procedure

The EMGM procedure (Supporting Material) was implemented in Matlab (The Mathworks). The initial values of the parameters that describe a mixture consisting of $K$ components were estimated by deleting a component from the previously estimated mixture consisting of $K$-1 components and adding two new components that are generated from the deleted one (Supporting Material). Additionally, one new Gaussian component is generated from the background component of the previously estimated mixture. This is done 3 times for each of the original $K$-1 Gaussian components and the background component, resulting in a total of $3K$ initializations. In the case of $K$=1, the initialization is done randomly 3 times. The procedure is stopped when the null hypothesis that the previously estimated $K$-1 component mixture is the correct one is fulfilled (Supporting Material). For this purpose, we simulated the distribution of likelihood increments when comparing the $K$-1 and $K$ component models under the null hypothesis. This distribution is obtained by simulating 100 datasets assuming the $K$-1 solution, and applying EMGM on each dataset, for both $K$-1 and $K$ mixture components. If the real likelihood increment has a $p$-value below 0.01 under the null hypothesis, it is assumed that the $K$-1 component solution is the correct one. Prior to analysis, the PALM data was split into overlapping 2×2 μm areas, and the EMGM analysis was performed on each area separately (Fig. S2 in the Supporting Material). Afterwards, identical mixture components in different EMGM results were combined according to a criterion based on the correlation between their posterior probabilities (Supporting Material).

# 3. Results

## 3.1 Expectation Maximization of a Gaussian Mixture (EMGM)

FAs display a substantial variety in size, shape, and density, and their substructure potentially as well. Quantifying the properties of the FA substructure with SMLM clustering methods is therefore challenging. Clusters in SMLM data are often characterized using the pair correlation function (26) or Ripley's $K(r)$ or $L(r)$ function (27). These functions describe the density around a certain point as a function of the distance $r$ from that point. As an illustration, we used PALM to image integrin β3 in a REF cell (Fig. 1A). We used Ripley's $L(r)$-$r$ function (28) to analyze a subset of the data (Fig. 1, B-D). This function shows a peak around 0.2 μm, indicating that the degree of clustering is highest on this length scale. However, it is difficult to interpret this result in terms of FA substructure properties, especially considering the heterogeneity in size and shape of the FAs themselves.

Such difficulties can be avoided by clustering methods that identify individual clusters based on criteria related to the local density of localizations, such as the nearest neighbor method (29) or density-based spatial clustering of applications with noise (DBSCAN) (30). We applied DBSCAN (31) to the same subset of the PALM data mentioned above (Fig. 1E). One value for the DBSCAN search radius identifies several substructures in the FA, while a larger value does not. However, the large search radius identifies two clusters that were considered to be background by the small search radius. It is clear that DBSCAN can handle the heterogeneity in size and shape of FAs, but identification of FA substructures largely depends on the values used for parameters that are related to a localization density threshold. Such a threshold is challenging to define, since FA substructures exhibit a variety of localization densities and can be closely packed (Fig. 1, A and B).

The difficulties related to established SMLM clustering methods prompted us to develop an approach based on EMGM (14). The main assumption of EMGM is that FAs can be modeled by a mixture of bivariate Gaussian probability distributions (Supporting Material). After choosing initial values for the parameters of each Gaussian component, the posterior probability that a certain localization was generated from a certain Gaussian component is evaluated (i.e. the expectation step). The Gaussian component parameters are then re-estimated using the new posterior probabilities (i.e. the maximization step) and the likelihood of the updated Gaussian mixture is calculated and checked for convergence.

In order to apply EMGM on SMLM data, we used a "greedy learning" approach (32) to initialize the parameters of the Gaussian components, and a model selection procedure based on hypothesis testing (33) to determine the number of components in the mixture (Supporting Material). However, the specific nature of SMLM data poses some additional challenges for EMGM. One problem is that not necessarily all localizations are part of the structure of interest, but can instead belong to a background. In case of a simple uniform background, the EMGM algorithm can readily be adjusted (Supporting Material). Moreover, the localizations in SMLM data contain measurement uncertainties (34). This localization uncertainty can be described by a spatial probability distribution that is usually modeled as a Gaussian. The EMGM can therefore be adapted by convolving the probability distributions that describe the mixture and the localization uncertainties (Supporting Material).

**3.2 Evaluation of EMGM on simulations**

The performance of the EMGM algorithm adapted for SMLM data was evaluated and validated by applying it to simulated data. We simulated mixtures consisting of $K$ closely spaced Gaussian components described by identical spatial probability distributions (i.e. 2D symmetric Gaussians with standard deviation $\sigma_x = \sigma_y = 20$ nm) and containing an identical number of positions (i.e. 100) (Fig. 2A, and Supporting Material). Such components have similar characteristics as nascent adhesions, or more speculatively, as the substructure of larger FAs.

First, we verified the performance of our proposed initialization scheme and model selection procedure. The results show that the simulated mixtures are correctly identified, provided $K$ is smaller than 10 (Fig. 2B, and Fig. S3 in the Supporting Material). We used $3K$ initializations for a mixture with $K$ components (Supporting Material). Increasing the number of initializations does not substantially improve the EMGM performance (Fig. S4 in the Supporting Material).

Next, we simulated the effect of a uniform localization background density $bg$ and a localization uncertainty $s$. The results indicate that the adapted EMGM correctly predicts $\sigma_{x,y}$ for values of $bg$ up to 25,000 #/µm$^2$ (Fig. 2C, and Fig. S5 in the Supporting Material). Our EMGM approach also captures the effect of the apparent increase in $\sigma_{x,y}$ due to localization uncertainties for values of $s$ up to 30 nm (Fig. 2D, Fig. S6 in the Supporting Material). Note that the largest values of $s$ and $bg$ included in these simulations are typically not encountered in "good quality" SMLM data.

The adapted EMGM should be able to distinguish closely spaced substructures inside FAs. Towards this end, we simulated Gaussian mixtures with a decreasing spacing $d_{x,y}$ between the component centers (Fig. 2E, and Supporting Material). The adapted EMGM performs well when $d_{x,y}$ is larger than 70 nm. A smaller $d_{x,y}$ results in a significant overlap in the spatial probability distribution of two adjacent components. Since one cannot assume that the substructures of FAs are radially symmetric, the component shape should also be accounted for by the EMGM algorithm. We simulated mixture components with decreasing $\sigma_x$ and simultaneously increasing $\sigma_y$ (Supporting Material). The results (Fig. 2F) clearly show that the algorithm correctly predicts the changing eccentricity $\sigma_x/\sigma_y$.

It should be noted that the results (Fig. 2) depend on the number of localizations that are contained by the components (Fig. S7 in the Supporting Material).

**3.3 Application of EMGM on experimental data**

To demonstrate the application of our EMGM algorithm, we made use of the SMLM data of a REF cell expressing mEos2-labelled integrin β3 (Fig. 1B). Similar to DBSCAN applied with the small search radius (Fig. 1E), EMGM also finds several FA substructures. Moreover, EMGM identifies two structures on the right as well, as indicated by the DBSCAN result using the large search radius (Fig. 1E).

We next proceeded to apply the EMGM algorithm on the whole PALM dataset (Fig. 1A). Since the simulation results (Fig. 2B) indicate that our algorithm works best for a small number of components, we reduce their number by applying a scanning procedure, consisting of splitting the original field of view into smaller overlapping areas, and by subsequently applying EMGM to each of these areas (Fig. S2). Afterwards, the results are combined, by merging identical Gaussian components in overlapping regions based on the correlation between their posterior probabilities, while excluding Gaussian components that belong to structures that were clipped during the splitting procedure (Supporting Material).

EMGM characterizes FA substructures in terms of bivariate Gaussian probability distributions. The properties of such a distribution can be translated into more intuitive properties using the error ellipse, i.e. the line of that describes a constant probability density. The major axis $a$ and the minor axis $b$ of the ellipse define its area and its shape (Fig. S8 in the Supporting Material). We therefore describe the FA substructure shape by the eccentricity $b/a$ (similar to the definition above). To calculate the area, we choose the $2\sigma$ error ellipse, corresponding to twice the standard deviation of the Gaussian component. This error ellipse defines the area in which there is a probability to find ~95% of all localizations belonging to the component. We pooled the area and eccentricity values of all identified mixture components in our PALM data set (Fig. 1G). Most components have an area below 0.5 $\mu m^2$ with a peak around 0.1 $\mu m^2$, and many exhibit some degree of eccentricity, with most values lower than 0.8. The EMGM algorithm also returns the posterior probability of each localization belonging to a specific Gaussian distribution, which gives the total number of localizations of each FA substructure (Supporting Material). Making the simplifying assumption that the localizations are uniformly distributed within the $2\sigma$ error ellipse, this leads to a characteristic localization density. Most FA substructures have a localization density below 2000 #/$\mu m^2$, and contain less than 100 localizations (Fig. 1G).

### 3.4 Integrin and paxillin

Following the evaluation of the adapted EMGM, we applied our method to investigate the substructure of FAs in cells growing on often-used fibronectin-coated substrates. We used PALM to image fixed REF cells ($n$ = 10) expressing paxillin or integrin β3 labelled with mEos2 (Fig. 3, A and B). To identify the FA substructure, we applied the adapted EMGM to each of these PALM datasets (Fig. 3C). As discussed above, the properties of individual mixture components, defined as bivariate Gaussians, can be described by three parameters: eccentricity, area, and number of localizations. We plotted these quantities as a function of each other, for both paxillin and integrin β3 (Fig. 3, D-F). Most mixture components contain between 10 and 100 localizations, and have an area between 0.01 and 1 $\mu m^2$ (Fig. 3D). The paxillin case displays a slightly more pronounced tail towards components that contain more localizations (up to 1000 localizations). When plotting the eccentricity as a function of the number of localizations (Fig. 3E), it is again apparent that the paxillin FA substructures can contain more localizations than the integrin ones. Furthermore, the mixture components in both cases appear to be eccentric, with most values below 0.7. The FA substructures containing less localizations appear to be somewhat more eccentric, a tendency that is more apparent in the paxillin case. A similar observation can be made when plotting the eccentricity as a function of the area (Fig. 3F). The larger the FA substructure, the more eccentric it seems to be. Interestingly, both paxillin and integrin objects seem to have similar areas, with a pronounced peak around 0.1 $\mu m^2$.

### 3.5 Nano-patterned substrates

The FA substructure properties (Fig. 3) have been obtained from REF cells growing on fibronectin-coated substrates. It can therefore not be guaranteed that the observed FA substructure is innate, it might simply be reflecting how the integrin binding sites on the fibronectin-coated substrate are organized on the nanoscale level. Such difficulties in interpretation of the data can be avoided by making use of a substrate where the integrin binding site locations are precisely controlled. We have therefore made use of block-copolymer micelle nanolithography (BCML) to pattern substrates with a quasi-hexagonal grid of 8 nm diameter gold nanoparticles (AuNPs) (18, 19) (Supporting Material). The AuNPs are functionalized with cyclic arginyl-glycyl-aspartic acid (cRGD) peptides, using a flexible polyethylene glycol (PEG) spacer. The area between the AuNPs is passivated with a PEG layer, ensuring that integrins can only adhere to the AuNPs. This enables a more unambiguous interpretation of the observed FA substructure. We chose a 56 nm spacing between the AuNPs, which was shown to result in normal cell adhesion (18). Furthermore, we also tested a 119 nm spacing, which poses more challenges for adhering cells (19).

We again imaged fixed REFs ($n$ = 10) expressing integrin β3 labelled with mEos2 (Fig. 4, A and B). The signal-to-noise ratio in the total internal reflection fluorescence (TIRF) images (Fig. 4A) appears to be lower than in the case of the fibronectin-coated substrates (Fig. 3A), suggesting that the grid of AuNPs affects the quality of the TIRF illumination. Next, we applied the adapted EMGM to each of the PALM datasets, in order to investigate the FA substructure (Fig. 4C). We plotted the number of localizations as a function of the area, for both the 56 and 119 nm AuNP spacings (Fig. 4, E and F). The fibronectin case (Fig. 4D) was added for comparison. It is clear that the objects on the fibronectin-coated substrate can contain up to 100 localizations, while the localization numbers on the 56 nm spacing substrate are generally below that level (Fig. 4, D and E). Interestingly, the FA substructure areas are very similar between both types of substrates, mostly between 0.01 and 1 μm² (Fig. 4, E and F). The FA substructure observed on the nano-patterned substrates does not appear in contradiction with the results obtained from fibronectin-coated substrates.

### 3.6 Isolated and overlapping mixture components

The interpretation of the EMGM results can be complicated (Fig. 3C, and Fig. 4C). Especially inside dense and large structures, which visually appear to be FAs, one can observe several components that overlap, based on their 2$\sigma$ error ellipses. The isolated mixture components, on the other hand, seem to correspond with smaller structures that could be nascent adhesions or focal complexes. We, therefore, performed a post-analysis step on EMGM results (Fig. 5A, and Supporting Material). We split the mixture components into two categories: the ones whose 1$\sigma$ error ellipse overlaps with at least one other 1$\sigma$ error ellipse, called the "overlapping" components, and the ones whose 1$\sigma$ error ellipse does not overlap with another one, called the "isolated" components. A new object can be calculated from a set of overlapping components, giving rise to a third category, called the "merged" components (Fig. 5A, and Supporting Material). Application of this merging procedure on a previously obtained EMGM result (Fig. 3C) shows that there are indeed several components that overlap (Fig. 5, B and C).

We applied the merging procedure on the EMGM results of REFs ($n$ = 10) expressing integrin β3 labelled with mEos2, growing on fibronectin-coated (Fig. 3D) and 56 nm spacing nano-patterned (Fig. 4D) substrates. As expected, on both types of substrate, the merged objects tend to have a larger area (up to 1 μm²) and contain more localizations (up to 1000 localizations) than the isolated and overlapping objects (Fig. 5D). The isolated components exhibit a similar behavior on both substrate types (Fig. 5E). Both cases exhibit FA substructures with an area between 0.01 and 0.1 μm², containing less than 100 localizations. The overlapping components are also not showing much difference between both substrate types, although the ones on the fibronectin-coated substrate can contain more localizations (Fig. 5F). Interestingly, the isolated and overlapping objects on the nano-patterned substrate also behave quite similarly (Fig. 5, E and F). The overlapping FA substructures are therefore not necessarily artifacts found by EMGM in a dense localization environment.

## 4. Discussion

We propose a new way to explore the properties of unknown structures as observed by SMLM. Using EMGM, we interpret patterns in SMLM data as a mixture of bivariate Gaussians. This approach allows to describe densely packed structures that can display strong heterogeneities in size, shape, and density, and is therefore well suited for investigation of the substructure of FAs.

However, application of EMGM to SMLM data is not without challenges. The result can be influenced by the choice of the initial values for the mixture component properties, and the number of components needs to be chosen as well. We identified an initialization procedure and a selection criterion for the number of components that gives good results for mixtures consisting of a small number of components (e.g. less than 10 for our simulated data). To allow analysis of larger numbers of components, we used a scanning procedure that essentially consists of splitting the SMLM data into smaller overlapping areas, and performing EMGM on each area separately. It is important to note that, unlike some SMLM clustering methods, the EMGM approach essentially does not depend on the choice of a free parameter (except for the area size of the scanning procedure).

The properties of SMLM data pose challenges to the classic EMGM algorithm. One complication is the localization uncertainty, which leads to an overestimation of the standard deviation of the Gaussian mixture components. An important contribution of this work is that we improved the EMGM approach to account for this effect. For "reasonable" localization uncertainties (e.g. below 30 nm for our simulated data) we found that the adapted EMGM worked well. We would like to point out that the effect of localization uncertainties is ignored by most existing SMLM clustering methods. Besides localization uncertainty, we also adjusted the EMGM algorithm to account for the presence of a uniform localization background. The method was found to perform excellently for any realistic level of background (e.g. up to 25,000 #/$\mu m^2$ for our simulated data).

To investigate the inner architecture of FAs, we performed SMLM imaging of FAs in fixed REF cells. We first explored the use of points accumulation in nanoscale topography (PAINT) (35) for imaging integrin β3 (Supporting Material). Our PAINT data suggests that that not all integrins are accessible for antibodies (Fig. S9 in the Supporting Material). To avoid antibody labeling problems, we therefore opted for PALM. We imaged integrin β3 and paxillin in fixed REF cells on fibronectin-coated substrates. The EMGM algorithm allowed us to identify integrin β3 objects with a typical area in the range between 0.01 and 1 $\mu m^2$, and containing between 10 and 100 localizations. Paxillin objects were found to have a similar area, but can contain more localizations, up to 1000. We attribute this difference to a tree-like organization of the FAs, rooting from isolated integrin islands, and expanding towards the actin filaments due to crosslinking and multivalent binding of paxillin and other proteins to their recruiting components. The equivalent diameter of the smallest objects was found to be around 100 nm (using the 2$\sigma$ error ellipse area, which is 0.01 $\mu m^2$ for the smaller objects). This indeed justifies the need for super-resolution microscopy to investigate the inner structure of FAs. Most objects were found to exhibit substantial eccentricity, with values down to 0.1. An algorithm that does not assume radial symmetry, such as EMGM, is therefore essential for the analysis of FA substructure.

A fibronectin coating is often used to ensure good cell adhesion to the substrate. However, it is important to rule out that the observed FA substructure is a mere artifact of the binding sites presented by such fibronectin-coated substrates. We therefore repeated the experiments on substrates that were patterned with a quasi-hexagonal grid of functionalized AuNPs. Our EMGM algorithm identified integrin β3 objects with areas in the same range as on fibronectin-coated substrates, while the number of localizations was lower, typically not exceeding 100. The FA substructure observed on the nano-patterned and fibronectin-coated substrates do not contradict each other.

The EMGM results sometimes display strongly overlapping mixture components, which is mathematically perfectly possible, but difficult to interpret. One possibility is that the background within the FAs is more complex than a simple uniform distribution. This could lead to the background partially being characterized by some of the mixture components, while the others are actual FA substructures. Note that our scanning procedure already captures background heterogeneities on the scale of the scanned areas. Another possibility is that a bivariate Gaussian is not the most accurate model for the FA subunits. To a certain extent, a post-analysis step can provide more insight. We performed a merging procedure that describes FA substructures either as an isolated Gaussian component, or a combination of several overlapping components. We hypothesize that the isolated components (areas between 0.001 and 0.01 $\mu m^2$, and number of localizations between 10 and 100) correspond to focal complexes or nascent adhesions. The overlapping mixture components, which appear to belong to FAs, have areas and localization numbers in the same range as the isolated components. This suggests that the observed objects are indicative of the real FA substructure. The merged components have a maximal area around 1 $\mu m^2$ and contain up to 1000 localizations, which can be interpreted as an upper limit for the FA substructure.

We envisage several ways in which our EMGM approach could be extended or adapted to allow a systematic and detailed study of the inner architecture of FAs. Several FA proteins could be investigated in multi-color mode to assess their spatial relationship. In this context, it could be of interest to develop an extension of EMGM that allows to investigate the "co-localization" of the mixture components. It would also be interesting to develop a 3D implementation of EMGM for the investigation of FA substructure in both the lateral and axial direction, as observed for instance by iPALM (17). It seems worthwhile to explore the possibility of incorporating other models than the Gaussian bivariate distribution, and other types of background besides the uniform one. Note that the effect of repetitive localizations on EMGM should be investigated, since photo-activatable fluorescent proteins can be localized more than once due to a phenomenon called "photoblinking" (36). Using transient transfection, a population of endogenous proteins will not fluorescently labeled, and the labelled proteins might be overexpressed. Techniques such as CRISPR/cas9 can bring solutions to this problem (37).

**5. Conclusion**

We have used PALM to investigate FAs in REF cells growing on fibronectin-coated substrates and specifically bio-functionalized nano-patterned substrates, on which ordered patterns of nanoscale adhesive spots were provided. To quantify the FA subunit properties, we developed a method based on EMGM that accounts for localization uncertainty and background. Analysis of our PALM data indicates that integrin β3 and paxillin structures within FAs have areas between 0.01 and 1 $\mu m^2$, contain 10 to 100 localizations, and can exhibit substantial eccentricities down to 0.1. We believe that our EMGM based approach is generic enough for the investigation of various other SMLM imaged nanoscale structures as well, especially for closely packed protein structures, or objects that display strong radial asymmetries and differences in size and density.

## Acknowledgements

The mEos2-paxillin-22 vector and the mEos2-Integrin-β3-N-18 vectors were kindly provided by Dr. Michael Davidson. H.D., J.P.S., and A.R. acknowledge the support of the Max Planck-EPFL Center for Molecular Nanoscience and Technology.

**Figures**

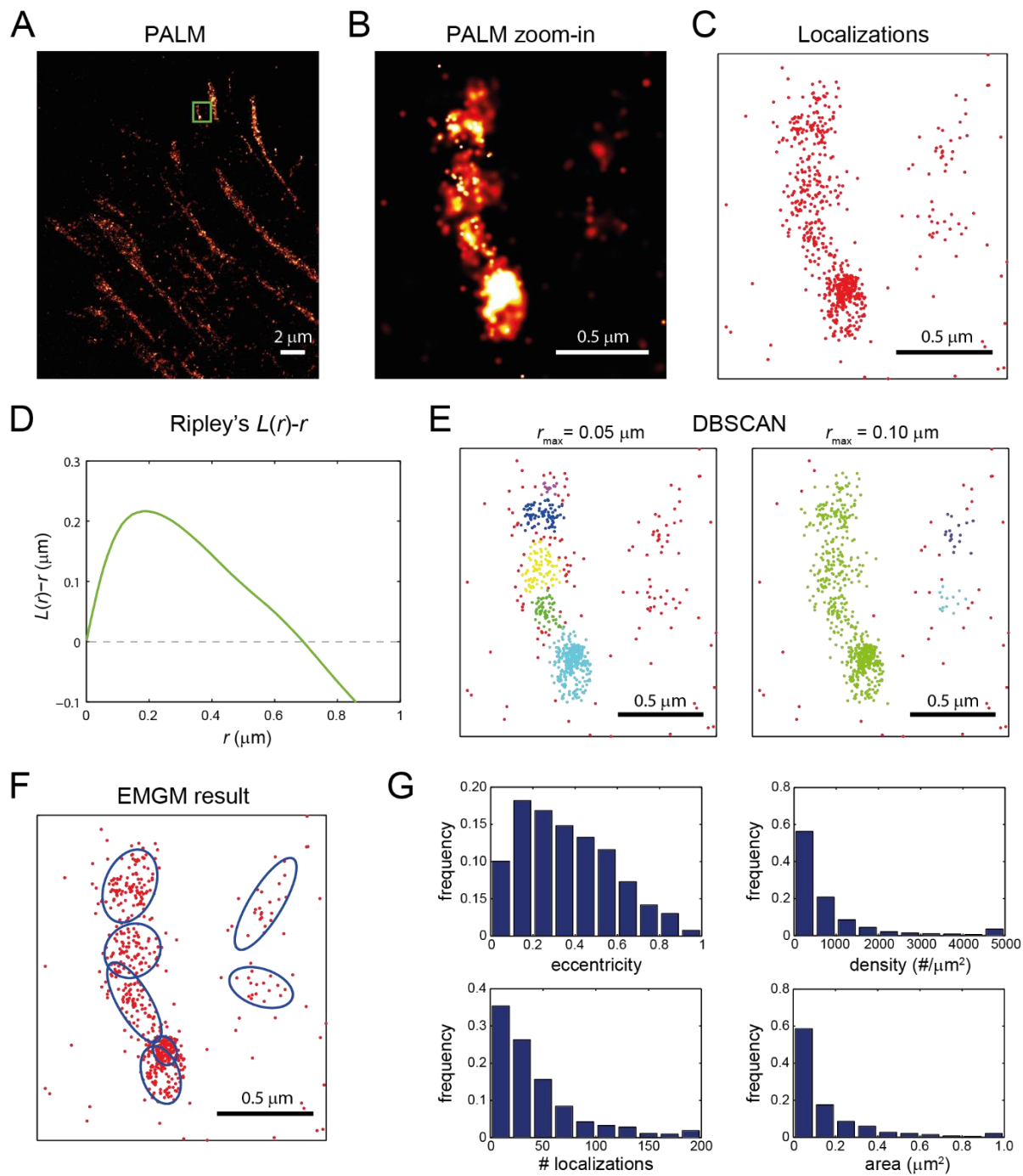

**Figure 1.** Application of SMLM clustering algorithms to PALM data of focal adhesions. (A) PALM image of a fixed REF cell expressing integrin β3 labelled with mEos2. (B) Zoom-in PALM image corresponding to the green rectangle in (A). (C) Scatter plot of the mEos2 localizations corresponding to the green rectangle in (A). (D) Ripley's $L(r)\text{-}r$ as a function of $r$, obtained from the localizations in (C). (E) Clusters obtained from the localizations in (C) by DBSCAN. The minimum number of localizations was set to 10, and two values were chosen for the maximum search radius $r_{max}$: 0.05 and 0.10 μm. The different colors of the localizations indicate to which cluster they belong, the background localizations are red. (F) Result of EMGM analysis of the localizations in (C). The red dots symbolize the localizations, and the blue ellipses the 2$\sigma$ error ellipses of the components. (G) Histograms showing the eccentricity $b/a$, localization density, number of localizations, and area $\pi ab$ of the 2$\sigma$ error ellipses of the components obtained by EMGM from the complete PALM data set in (A). The rightmost bins contains all values within that bin and larger.

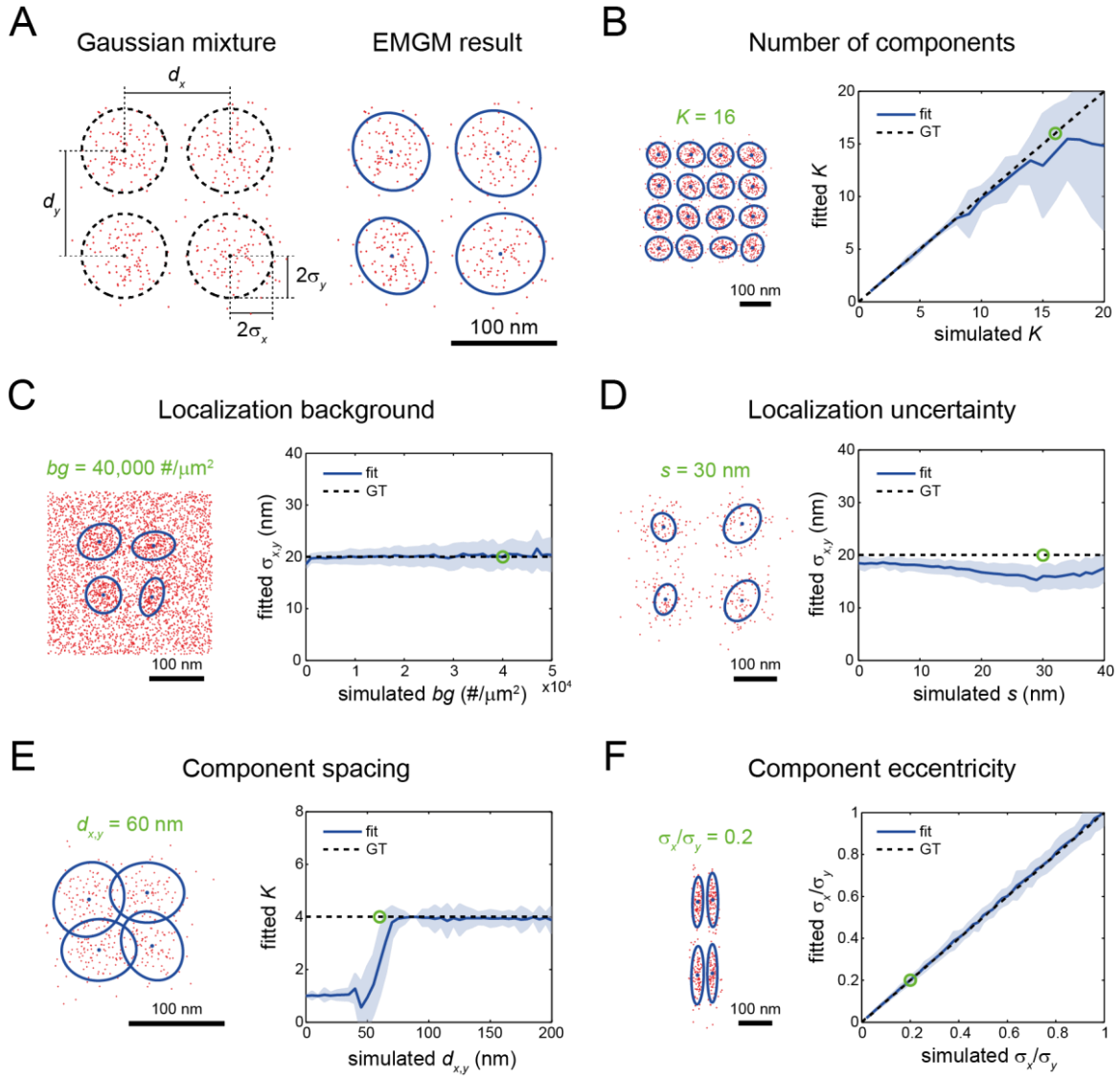

**Figure 2.** Evaluation of EMGM using simulated data. (A) On the left, an example of a simulated Gaussian mixture consisting of $K = 4$ components, each containing 100 localizations, described by a symmetric 2D Gaussian distribution with a standard deviation $\sigma_x = \sigma_y = 20$ nm. The Gaussian centers are placed in a square grid with spacing $d_{x,y} = 100$ nm. On the right, the EMGM result. The red dots symbolize the localizations. The blue dots symbolize the center positions and the blue ellipses the $2\sigma$ error ellipses of the components. (B) On the right, the average number of mixture components correctly identified by EMGM as a function of the simulated $K$. On the left, an example EMGM result for $K = 16$. (C) On the right, the average standard deviation $\sigma_{x,y}$ of the mixture components calculated by EMGM as a function of the simulated localization background density $bg$. On the left, an example EMGM result for $bg = 40,000$ #/$\mu$m². (D) On the right, the average $\sigma_{x,y}$ calculated by EMGM as a function of the simulated localization uncertainty $s$. On the left, an example EMGM result for $s = 30$ nm. (E) On the right, the average number of mixture components correctly identified by EMGM as a function of the simulated spacing $d_{x,y}$. On the left, an example EMGM result for $d_{x,y} = 60$ nm. (F) On the right, the average eccentricity $\sigma_x/\sigma_y$ of the mixture components calculated by EMGM as a function of the simulated $\sigma_x/\sigma_y$. On the left, an example EMGM result for $\sigma_x/\sigma_y = 0.2$. The simulated Gaussian mixtures in (C-F) consist of $K = 4$ components, similar to (A). The dashed lines in (B-F) represent the ground truth (GT) and the shaded areas the standard deviation ($n = 100$).

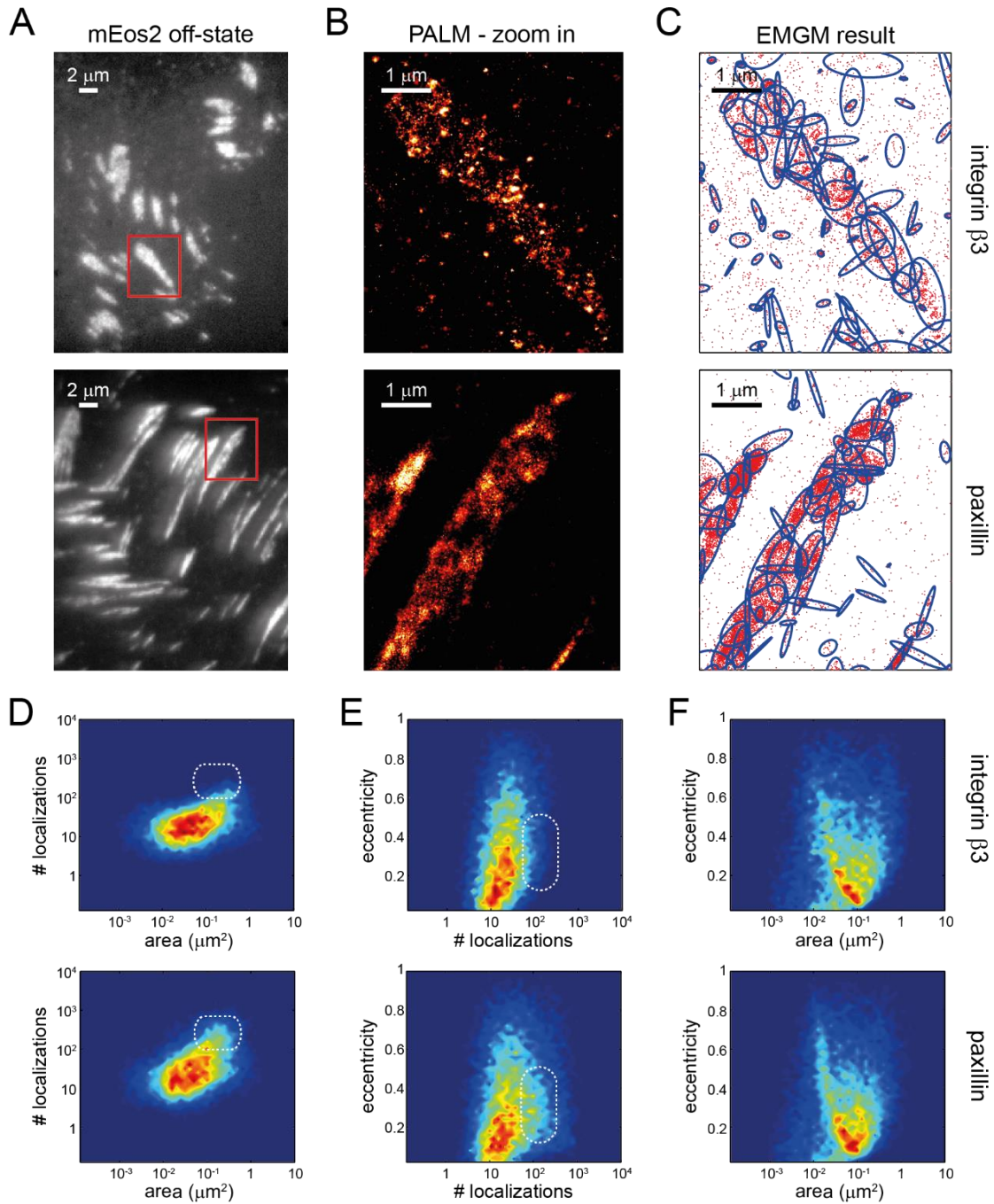

**Figure 3.** EMGM analysis of PALM data of integrin β3 or paxillin on fibronectin-coated substrates. (A) Summed TIRF images of the mEos2 off-state of fixed REF cells expressing integrin β3 or paxillin labelled with mEos2, growing on fibronectin-coated substrates. (B) Zoom-in PALM images corresponding to the green rectangles in (A). (C) Result of the EMGM analysis of the PALM data shown in (B). The red dots symbolize the localizations, and the blue ellipses the 2σ error ellipses of the mixture components. (D-F) Result of the EMGM analysis of PALM data corresponding to different REF cells (*n* = 10): (D) number of localizations in each mixture component as a function of the area of its 2σ error ellipse, (E) eccentricity of the 2σ error ellipse of each mixture component as a function of its number of localizations, (F) eccentricity of the 2σ error ellipse of each mixture component as a function of its area. The dashed white rounded rectangles in (D) and (E) are visual guides.

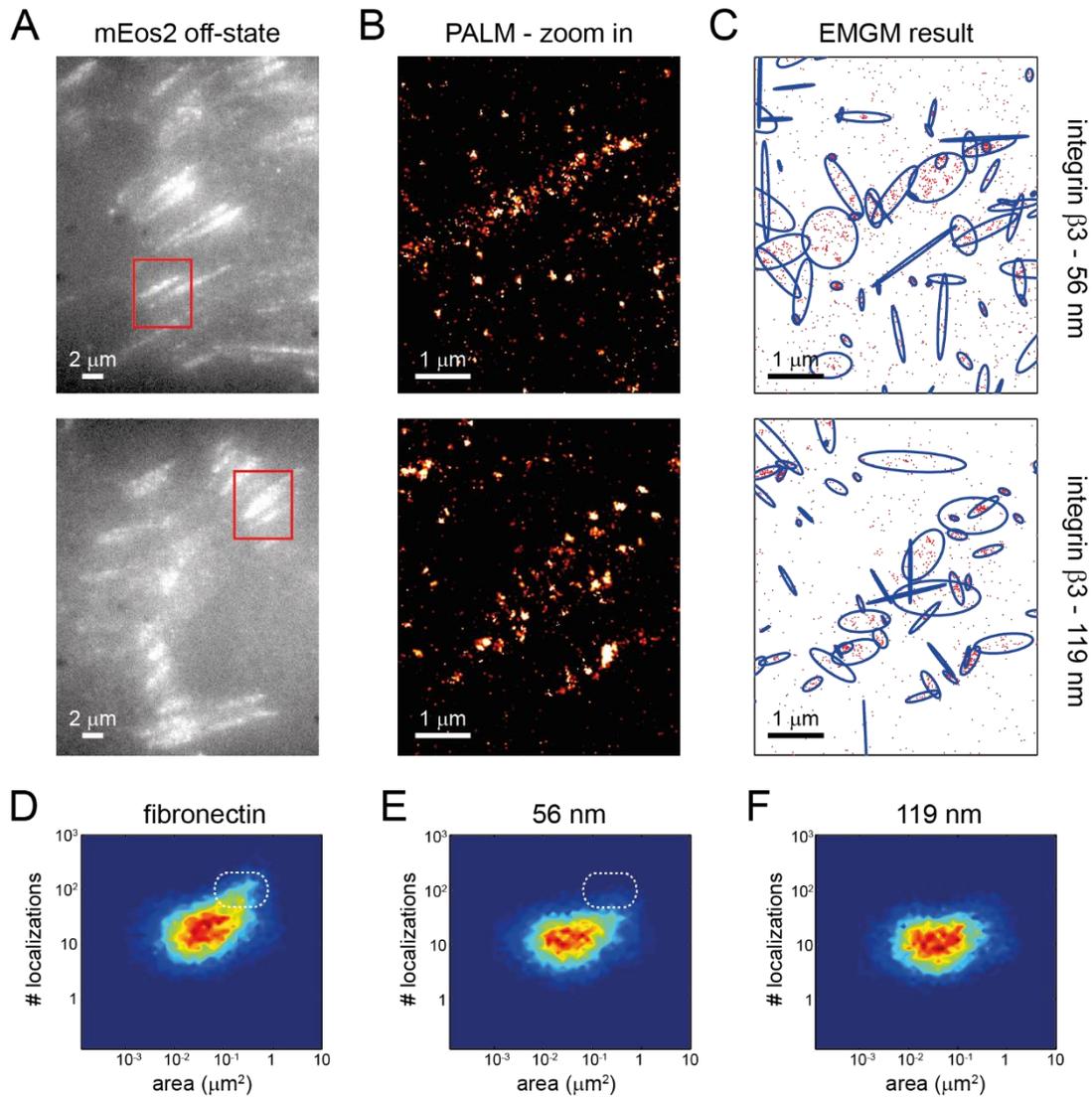

**Figure 4.** EMGM analysis of PALM data of integrin β3 on nano-patterned substrates. (A) Summed TIRF images of the mEos2 off-state of fixed REF cells expressing integrin β3 labelled with mEos2, growing on nano-patterned substrates with 56 nm or 119 nm spacing between the AuNPs. (B) Zoom-in PALM images corresponding to the green rectangles in (A). (C) Result of the EMGM analysis of the PALM data shown in (B). The red dots symbolize the localizations, and the blue ellipses the $2\sigma$ error ellipses of the mixture components. (D-F) Result of the EMGM analysis of PALM data corresponding to different REF cells ($n$ = 10). The number of localizations in each mixture component is shown as a function of the area of its $2\sigma$ error ellipse, for (D) fibronectin coated substrates (Fig. 3D), (E) nano-patterned substrates with 56 nm spacing, (F) nano-patterned substrates with 119 nm spacing. The dashed white rounded rectangles in (D) and (E) are visual guides.

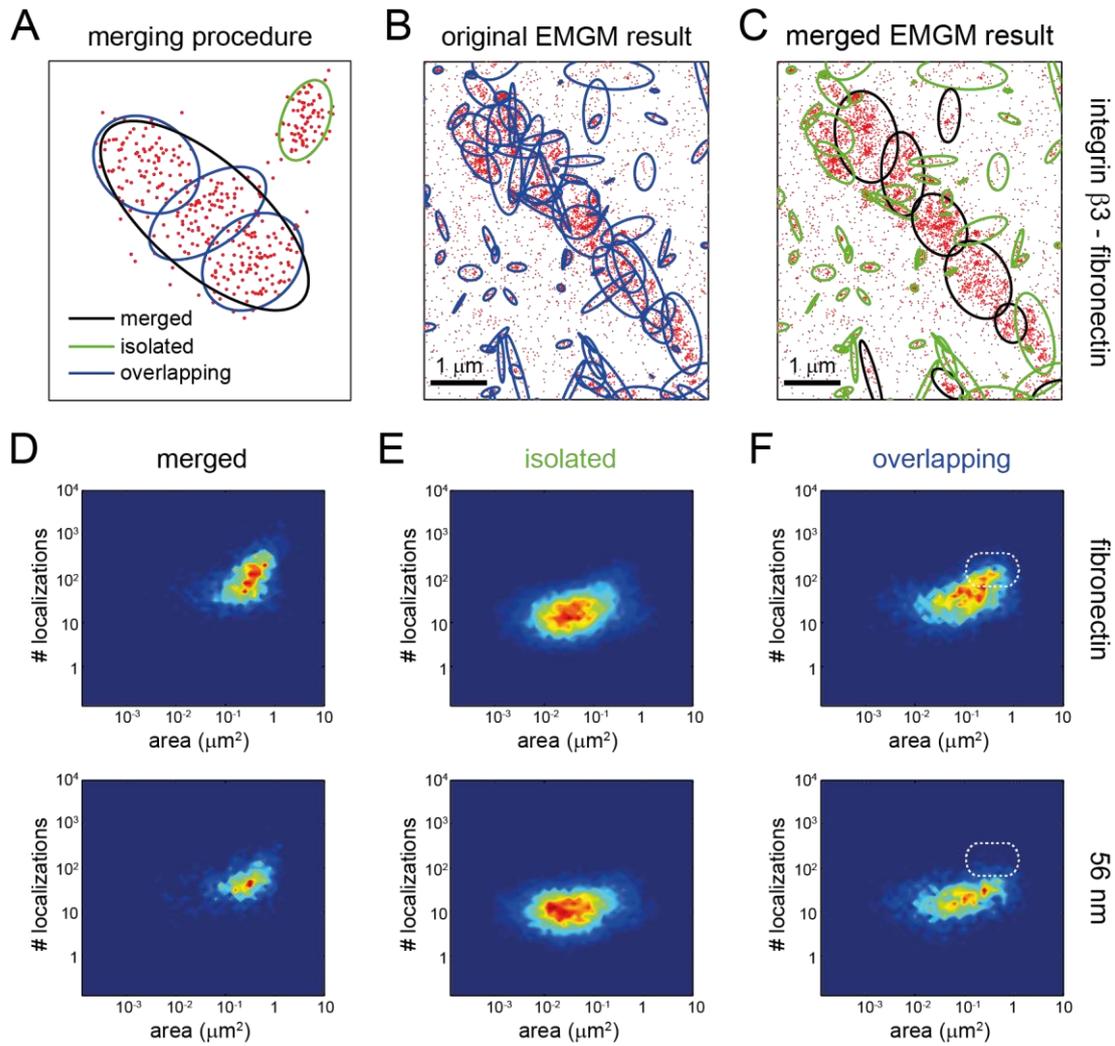

**Figure 5.** Merging procedure applied on EMGM results for integrin β3. (A) Illustration of the concept of merging overlapping mixture components based on overlapping error ellipses. The red dots symbolize the localizations. The black/green/blue ellipses represent the $2\sigma$ error ellipses of the merged/isolated/overlapping mixture components. (B) EMGM result for PALM data of a fixed REF cell growing on a fibronectin-coated substrate expressing integrin β3 labelled with mEos2 (Fig. 3C). (C) Result of the merging procedure applied on the EMGM result in (B). (D-F) Result of the merging procedure applied on EMGM results for integrin β3 (Fig. 4D, and Fig. 5D). The number of localizations in each mixture component is shown as a function of the area of its $2\sigma$ error ellipse, for (D) the merged components, (E) the isolated components, and (F) the overlapping components. The dashed white rounded rectangles in (F) are visual guides.

# Supporting Material

# Supporting Text

## 1. Expectation maximization of a Gaussian mixture (EMGM)

### 1.1 Classic algorithm

We apply expectation maximization of a Gaussian mixture (EMGM) [1] on single-molecule localization (SMLM) data to investigate the substructure of focal adhesions (FAs). The main assumption is that the FA subunits can be described as bivariate Gaussians. The spatial probability distribution of an FA subunit is thus given by:

$$N(\boldsymbol{r}|\boldsymbol{\mu},\boldsymbol{\Sigma}) = \frac{1}{2\pi\sqrt{|\boldsymbol{\Sigma}|}}\exp\left(-\frac{1}{2}(\boldsymbol{r}-\boldsymbol{\mu})^{\mathrm{T}}\cdot\boldsymbol{\Sigma}^{-1}\cdot(\boldsymbol{r}-\boldsymbol{\mu})\right) \tag{1}$$

where $\boldsymbol{r}$ is the position in which the Gaussian is being evaluated, $\boldsymbol{\mu}$ the center position of the Gaussian, and $\boldsymbol{\Sigma}$ the covariance matrix of the Gaussian. Assume one or more FAs consisting out of $N$ positions $\boldsymbol{r}_n$. According to our assumption, these FAs can be modeled by a mixture of bivariate Gaussians. Assume that that this mixture consists of $K$ components with the weight of component $k$ described by the mixing coefficient $\pi_k$. These mixing coefficients fulfil the condition:

$$\sum_{k=1}^{K}\pi_k = 1 \tag{2}$$

Expectation maximization is a popular algorithm to identify the properties $\boldsymbol{\mu}_k$, $\boldsymbol{\Sigma}_k$ and $\pi_k$ of each component the Gaussian mixture. After choosing initial values, the expectation step consists of evaluating the posterior probability that localization $\boldsymbol{r}_n$ was generated from component $k$:

$$\gamma_{nk} = \frac{\pi_k N(\boldsymbol{r}_n|\boldsymbol{\mu}_k,\boldsymbol{\Sigma}_k)}{\sum_{j=1}^{K}\pi_j N(\boldsymbol{r}_n|\boldsymbol{\mu}_j,\boldsymbol{\Sigma}_j)} \tag{3}$$

In the maximization step, the parameters are re-estimated using the posterior probabilities:

$$\boldsymbol{\mu}_k^{\mathrm{new}} = \frac{1}{N_k}\sum_{n=1}^{N}\gamma_{nk}\boldsymbol{r}_n$$

$$\boldsymbol{\Sigma}_k^{\mathrm{new}} = \frac{1}{N_k}\sum_{n=1}^{N}\gamma_{nk}(\boldsymbol{r}_n-\boldsymbol{\mu}_k^{\mathrm{new}})\cdot(\boldsymbol{r}_n-\boldsymbol{\mu}_k^{\mathrm{new}})^{\mathrm{T}} \tag{4}$$

$$\pi_k^{\mathrm{new}} = \frac{N_k}{N}$$

Where $N_k$ is defined as the number of localizations that belong to component $k$:

$$N_k = \sum_{n=1}^{N}\gamma_{nk} \tag{5}$$

Finally, the likelihood of the updated Gaussian mixture is calculated and checked for convergence:

$$\mathcal{L} = \prod_{n=1}^{N}\sum_{j=1}^{K}\pi_k^{\mathrm{new}}N(\boldsymbol{r}_n|\boldsymbol{\mu}_j^{\mathrm{new}},\boldsymbol{\Sigma}_j^{\mathrm{new}}) \tag{6}$$

If the convergence criterion is not satisfied, the expectation and maximization steps described in Eqs. (3) and (4) are repeated.

### 1.2 Initialization by greedy learning

EMGM is known to be sensitive to local maxima. To avoid finding such a solution, initial values of the parameters $\boldsymbol{\mu}_k$, $\boldsymbol{\Sigma}_k$ and $\pi_k$ (see Supporting Text, Section 1.1) need to be chosen sufficiently close to the real values. In the context of SMLM, these values are not known. Although several approaches have been reported in order to initialize the model parameters for EMGM, there is no widely accepted method. Popular approaches are randomly generating the initial parameter values, or estimating them using the $k$-means clustering algorithm [1].

An interesting alternative to these initialization methods is the so-called "greedy learning" approach [2], based on repeating the EMGM by starting from a trivial Gaussian mixture consisting of one component, and each time adding an extra component. The EMGM solution obtained for a $P$-1 component mixture is used as initialization for the $P$ component mixture, by deleting one component and inserting two random components, based on the deleted one. This can be done $P$-1 times, for each component of the old mixture, and the solution with the highest likelihood is retained. By doing so, one proceeds until a desired number of components $K$ is attained. Additionally, each step consisting of $P$-1 initializations can be repeated $Q$ times to increase the accuracy of the result. The total number of EMGM repeats to obtain the correct solution of $K$ components is thus given by $Q\left(1 + \sum_{i=1}^{K} i\right)$.

This shows that the initialization procedure becomes computationally more expensive for datasets containing more components. The computation time on a mid-range personal computer for the simulations shown in Fig. 2 ranged from ~3 s (for $K = 1$ and $Q = 3$) to ~1000 s (for $K = 20$ and $Q = 3$). Note that we actually used $Q\left(1 + \sum_{i=1}^{K}[i + 1]\right)$ initializations due to an extra background "component", see Supporting Text, Section 1.4.

## 1.3 Model selection by hypothesis testing

When applying EMGM, the number of components $K$ for the Gaussian mixture needs to be chosen. In the context of SMLM, this number is unknown. In order to select the most appropriate number of components, one can repeat the EMGM procedure for a range of $K$ values. The likelihood value is not a good selection criterion, as increasing the number of components increases the likelihood monotonously. A solution provided by information theory is the Akaike or Bayes information criterion [1], which penalizes an increasing number of components and therefore leads to a maximum value for a certain $K$ value. However, this value has been reported to typically overestimate the real number of components [3].

Hypothesis testing can provide a more conservative approach towards selecting to right mixture model [4]. Assume two mixtures calculated by EMGM, one containing $K$-1 components and the other containing $K$ components. The $K$ component model will have a larger likelihood than the $K$-1 component model. Consider the null hypothesis that the $K$-1 component model is the correct one, which will correspond to a specific distribution of likelihood increments. If the real model consists of more than $K$-1 components, the likelihood increment can be expected to be larger than the values described by the null hypothesis distribution. This distribution, however, is unknown, but can be simulated from the identified $K$-1 component model, i.e. a number of bootstrapped data sets are generated assuming the null hypothesis and the increments in likelihood are obtained by applying EMGM for both $K$-1 and $K$ components. Comparing the real likelihood increment with the bootstrap null hypothesis distribution allows to determine the *p*-value, in turn allowing to accept or reject the null hypothesis. Choosing the maximum allowed *p*-value sufficiently small, e.g. equal to 0.01, means that there is only a 1% chance to select a mixture model that contains too many components, preventing overestimation of the number of components.

## 1.4 Localization background

While initialization and model selection issues are inherent to EMGM, other problems arise because of the nature of SMLM data. One important problem is that not necessarily all localizations are part of FAs, but instead can belong to a background. Consider an SMLM dataset consisting of $N$ positions that belong to a mixture of multivariate Gaussians, and an extra $N_b$ positions that belong to a background, within an area $A$. In case of a simple uniform background, the probability distribution of the background localizations is given by:

$$B = \frac{1}{A} \tag{7}$$

The algorithm can readily be adjusted to incorporate the background described by $B$. First of all, the posterior probability that localization $\boldsymbol{r}_n$ was generated from component $k$ (see Eq. (3)) is now given by:

$$\gamma_{nk} = \frac{\pi_k N(\boldsymbol{r}_n | \boldsymbol{\mu}_k, \boldsymbol{\Sigma}_k)}{\sum_{j=1}^{K} \pi_j N(\boldsymbol{r}_n | \boldsymbol{\mu}_j, \boldsymbol{\Sigma}_j) + B} \tag{8}$$

And an equivalent posterior probability for the background can be defined as:

$$\delta_n = \frac{B}{\sum_{j=1}^{K} \pi_j N(\boldsymbol{r}_n | \boldsymbol{\mu}_j, \boldsymbol{\Sigma}_j) + B} \tag{9}$$

The re-estimation of the parameters $\boldsymbol{\mu}_k$ and $\boldsymbol{\Sigma}_k$ can be done as before, while the re-estimation of the mixing coefficients (see Eq. (4)) has to be adjusted as follows:

$$\pi_k^{\text{new}} = \frac{N_k}{N + N_b} \tag{10}$$

where $N_b$ can be calculated using the background posterior probabilities:

$$N_b = \sum_{n=1}^{N} \delta_n \tag{11}$$

Finally, the calculation of the likelihood of the updated Gaussian mixture (see Eq. (6)) is adjusted as follows:

$$\mathcal{L} = \prod_{n=1}^{N+N_b} \left\{ \sum_{j=1}^{K} \pi_k^{\text{new}} N(\boldsymbol{r}_n | \boldsymbol{\mu}_j^{\text{new}}, \boldsymbol{\Sigma}_j^{\text{new}}) + B \right\} \tag{12}$$

The background can effectively be considered as an extra component of the Gaussian mixture, requiring an adaptation of the initialization procedure described in Supporting Text, Section 1.2. Initialization of a $P$ component Gaussian mixture is done $P$ times instead of $P - 1$ times (i.e. $P - 1$ initializations corresponding to each component of the previous solution, and 1 initialization corresponding to the background of the previous solution).

## 1.5 Localization uncertainty

The localizations in SMLM data contain measurement uncertainties [5]. The localization uncertainty can be described as an extra contribution $\boldsymbol{\varepsilon}$ to the real position of the molecule. This contribution is described by a spatial probability distribution that is usually modeled as a Gaussian:

$$E(\boldsymbol{\varepsilon} | s) = \frac{1}{2\pi s} \exp\left(-\frac{|\boldsymbol{\varepsilon}|^2}{2s^2}\right) \tag{13}$$

The standard deviation $s$ is often termed as the localization uncertainty or precision. An observed localization $\boldsymbol{r}$ belonging to component $k$ is described by the sum of $\boldsymbol{\varepsilon}$ and the real emitter position. Since both variables are independent, the spatial probability distribution of their sum is given by the convolution of their corresponding spatial probability distributions (see Eqs. (1) and (13)):

$$N(\boldsymbol{r}|\boldsymbol{\mu}_k, \boldsymbol{\Sigma}_k, s) = \int\limits_{-\infty}^{+\infty} E(\boldsymbol{r} - \boldsymbol{r}'|s) N(\boldsymbol{r}'|\boldsymbol{\mu}_k, \boldsymbol{\Sigma}_k) \, d\boldsymbol{r}' \tag{14}$$

This is the convolution of two bivariate Gaussians, which can be solved as [6]:

$$N(\boldsymbol{r}|\boldsymbol{\mu}_k, \boldsymbol{\Sigma}_k, s) = \frac{1}{2\pi\sqrt{|\boldsymbol{\Sigma}_k + s^2\boldsymbol{I}|}} \exp\left(-\frac{1}{2}(\boldsymbol{r} - \boldsymbol{\mu}_k)^{\mathrm{T}} \cdot (\boldsymbol{\Sigma}_k + s^2\boldsymbol{I})^{-1} \cdot (\boldsymbol{r} - \boldsymbol{\mu}_k)\right) \tag{15}$$

where $\boldsymbol{I}$ is the identity matrix. This expression describes the observed spatial probability distribution of component $k$. In order to incorporate the effect of the localization uncertainty in EMGM, we need to adjust the algorithm in two ways. First of all, the expectation step needs to be adjusted, since the expression for the posterior probability $\gamma_{nk}$ of molecule position $\boldsymbol{r}_n$ of component $k$ contains the spatial probability distribution of that component (see Eq. (3)). Substitution of Eq. (15) in Eq. (3) yields the adjusted posterior probability:

$$\gamma_{nk} = \frac{\pi_k N(\boldsymbol{r}_n|\boldsymbol{\mu}_k, \boldsymbol{\Sigma}_k, s_n)}{\sum_{j=1}^{K} \pi_j N(\boldsymbol{r}_n|\boldsymbol{\mu}_j, \boldsymbol{\Sigma}_j, s_n)} \tag{16}$$

where $s_n$ is the localization uncertainty corresponding to localization $\boldsymbol{r}_n$. Secondly, the maximization step needs to be adjusted, because the apparent spatial probability distribution is a bivariate Gaussian with a covariance matrix equal to $\boldsymbol{\Sigma}_k + s^2\boldsymbol{I}$ (see Eq. (15)). This means that the presence of localization uncertainties affects both the shape and size of the observed component $k$. The re-estimation of the covariance matrix (see Eq. (4)) should be adjusted as follows:

$$\boldsymbol{\Sigma}_k^{\mathrm{new}} = \frac{1}{N_k} \sum_{n=1}^{N} \gamma_{nk} \{(\boldsymbol{r}_n - \boldsymbol{\mu}_k^{\mathrm{new}}) \cdot (\boldsymbol{r}_n - \boldsymbol{\mu}_k^{\mathrm{new}})^{\mathrm{T}} - s_n^2\boldsymbol{I}\} \tag{17}$$

The contribution coming from the localization uncertainty is included within the sum, since the value of the localization uncertainty can change for different localizations. Note that Eq. (17) suggests that the covariance matrix values of certain mixture components can possibly become negative during the EMGM procedure. If this occurs during EMGM, the covariance matrix is not updated, and the value of the previous iteration is retained.

## 2. Simulations

### 2.1 Simulation details

The simulations shown in Fig. 2 were performed in Matlab (The Mathworks). Briefly, Gaussian mixtures consisting of $K$ components were simulated. The localizations in each component were obtained from a Gaussian probability distribution, using the Matlab function *mvnrnd*. The Gaussian standard deviation was $\sigma_x = \sigma_y$ = 20 nm (except for Fig. 2F), and the number of localizations for each component was $N_k$ = 100. The number of mixture components $K$ was varied between 1 and 20 in Fig. 2B, and fixed at 4 in Fig. 2C-F. The centers of the mixture components were placed in a square grid with a spacing $d_{x,y}$ equal to five times $\sigma_{x,y}$ (except for Fig. 2D).

A uniform localization background was added in Fig. 2C by randomly generating a number of localizations from a uniform distribution, using the Matlab function *rand*. The number of background localizations was determined from the localization background density $bg$, which was varied between 0 and 50,000 #/$\mu m^2$, in steps of 1000 #/$\mu m^2$. The effect of the localization uncertainty shown in Fig. 2D was simulated by adding to each localization coordinate a value randomly generated from a Gaussian distribution with standard deviation $s$, using the Matlab function *randn*. The value of the localization uncertainty $s$ was varied from 0 to a 40 nm, in steps of 1 nm. To account for the apparent increase in component size, the spacing between the component centers was adjusted to five times $\sqrt{\sigma_{x,y}^2 + s^2}$. In Fig. 2E, the spacing $d_{x,y}$ between the component centers was increased from 0 to 200 nm, in steps of 5 nm. The changing component eccentricity shown in Fig. 2F was simulated by increasing the component standard deviation $\sigma_x$ from 2.8 to 20 nm, and simultaneously decreasing the standard deviation $\sigma_y$ from 140 to 20 nm, resulting in eccentricities $\sigma_x/\sigma_y$ increasing from 0.02 to 1. For each case, 100 simulations were performed.

### 2.2 Number of mixture components

The simulation results in Fig. 2B show that EMGM increasingly underestimates the number of mixture components for an increasing value of $K$. Additionally, the number of non-existing components (i.e. false positives) identified by EMGM also increases with $K$, as illustrated in Fig. S3, A and B. We define $K_{id}$ as the number of mixture components correctly identified by EMGM, and $K_{fp}$ as the number of false positive components found by EMGM. Using the simulated data from Fig. 2B, we calculated the probability of obtaining a completely correct EMGM result (i.e. $K_{id} = K$ and $K_{fp}$ = 0) as a function of $K$. The results are shown in Fig. S3C. For mixtures with $K$ < 10, this probability is on average equal to 94%. For larger numbers, the method starts to underestimate $K$, most likely because the contribution of correctly fitting individual components to the total likelihood becomes smaller with an increasing number. Fig. S3D shows the average values of $K_{id}$ and $K_{fp}$ as a function of $K$. The average number of false positives is smaller than 1 for mixtures with $K$ < 10.

### 2.3 Number of initializations

The initialization procedure described in Supporting Text, Section 1.2 consists of $P$-1 separate initializations for a $P$ component Gaussian mixture. If the localization background is considered as an extra component, the procedure actually consists of $P$ separate initializations for a $P$ component mixture (see Supporting Text, Section 1.4). This procedure can be repeated several times $Q$ to improve the accuracy of the EMGM result, resulting in a total of $QP$ initializations for a $P$ component Gaussian mixture. In order to investigate the effect of the value of $Q$ on the EMGM performance, we performed simulations similar to the ones shown in Fig. 2B, for different values of $Q$. Fig. S4A shows that an increasing $Q$ results in less underestimation of $K$, although the improvement is small for $Q > 3$. The number of false positive components $K_{fp}$ does not seem to be affected by the value of $Q$, see Fig. S4B.

## 2.4 Localization background

The adapted EMGM performs excellently in the presence of a uniform localization background (see Fig. 2C and Fig. S5, A and B). Only for values of the localization background density that are not representative for our experimental conditions (e.g. $bg$ = 50,000 #/µm$^2$ in Fig S5C), the algorithm starts to underestimate the true amount of mixture components and to find false positive components. Using the simulated data from Fig. 2C, we calculated the probability of obtaining a completely correct EMGM result (i.e. $K_{id} = K$ and $K_{fp} = 0$) as a function of $bg$ (see Fig. S5C). For mixtures with $bg < 25,000$ #/µm$^2$, this probability is on average equal to 93%. Fig. S5D shows the average values of $K_{id}$ and $K_{fp}$ as a function of $bg$, confirming that the EMGM performance deteriorates for values larger than 25,000 #/µm$^2$. This is not a surprise, since the characteristic localization density of the component mixtures themselves is lower (each component counts 100 localization and has a standard deviation of $\sigma_{x,y}$ = 20 nm, resulting in a $2\sigma$ ellipse area of 0.016 µm$^2$, which yields a characteristic localization density around 20,000 #/µm$^2$).

## 2.5 Localization uncertainty

The simulation results in Fig. 2D show that the estimated standard deviation $\sigma_{x,y}$ of the mixture components is slightly affected by an increasing localization uncertainty $s$. However, as illustrated in Fig. S6, A-C, an increasing value of $s$ can have an important impact on the values of $K_{id}$ and $K_{fp}$. We assessed the probability of obtaining a completely correct EMGM result (i.e. $K_{id} = K$ and $K_{fp} = 0$) as a function of $s$, using the simulated data shown in Fig. 2D. The results are shown in Fig. S6D, indicating that the probability decreases strongly when $s$ becomes larger than 30 nm. This is to be expected, since the localization uncertainty is larger than the standard deviation $\sigma_{x,y}$ = 20 nm of the mixture components itself. Fig. S6E shows $K_{id}$ and $K_{fp}$ as a function of $s$. For localization uncertainties larger than 30 nm, the average number of correctly identified components slightly decreases, while the average number of false positives increases more strongly.

## 2.6 Number of localizations

The simulations presented in Fig. 2 describe Gaussian mixtures with components that each consist of $N_k$ = 100 localizations. However, as illustrated in Fig. S7, A-C, the performance of the EMGM algorithm can depend on the value of $N_k$. We assessed the probability of obtaining a completely correct EMGM result (i.e. $K_{id} = K$ and $K_{fp} = 0$) as a function of $N_k$, using simulations similar to Fig. 2A. The results are shown in Fig. S7D, indicating that the probability decreases strongly when $N_k$ becomes smaller than 50. Fig. S7E shows $K_{id}$ and $K_{fp}$ as a function of $N_k$, indicating that this low probability is mainly due to EMGM not detecting all mixture components for such low number of localizations.

# 3. Applying EMGM on experimental data

## 3.1 Scanning procedure

The number of FA substructures present in a typical SMLM dataset is not known, and can be assumed to be larger than 10. However, the simulation results in Fig. 2B indicate that the EMGM analysis is optimal when the Gaussian mixture consists of a smaller number of components. We therefore split the SMLM dataset into smaller subsets and perform the EMGM analysis on each subset separately. This can be done simply by scanning the original region of interest along non-overlapping square subregions with side length $L$, as illustrated in Fig. S2, A and B. However, this scanning procedure clips Gaussian mixture components that are not completely contained in a single subregion. A solution is repeating the scan with subregions that are shifted over a distance equal to $L/2$. If this shift is done in three different directions, as shown in Fig. S2, B-E, each component with dimensions below $L/2$ is completely included in at least one subregion of at least one scan. Considering that the FA substructures of interest have sizes below the diffraction limit, we choose $L = 2$ μm.

## 3.2 Combining procedure

Combining the EMGM results obtained from the scanning procedure described in Supporting Text, Section 3.1 consists of two steps: (1) the EMGM results of the subregions within each separate scan need to be combined, resulting in four different EMGM descriptions of the same original dataset, and (2) combining these four results yields the final EMGM result.

For the first step, we make the approximation that all components identified in a subregion are completely described by the localizations within that subregion. The posterior probability (see Eq. (3)) of a localization within a certain subregion belonging to a component identified in another subregion will therefore be zero. This means that the posterior probabilities of all $M$ subregions of a single scan can be assembled into a sparse matrix $\boldsymbol{\gamma}_{\text{scan}}$ to describe the posterior probabilities of the full dataset:

$$\boldsymbol{\gamma}_{\text{scan}} = \begin{bmatrix} \boldsymbol{\gamma}_1 & \cdots & \mathbf{0} \\ \vdots & \ddots & \vdots \\ \mathbf{0} & \cdots & \boldsymbol{\gamma}_M \end{bmatrix} \tag{18}$$

where the matrices $\boldsymbol{\gamma}_i$ describe the posterior probabilities of the localizations inside subregion $i$, with $i = 1, \ldots, M$. The posterior probabilities corresponding to the full dataset for a localization to belong to the background (see Eq. (9)) are similarly given by:

$$\boldsymbol{\delta}_{\text{scan}} = \begin{bmatrix} \boldsymbol{\delta}_1 \\ \vdots \\ \boldsymbol{\delta}_M \end{bmatrix} \tag{19}$$

This approximation is not optimal for the case of Gaussian mixture components being clipped, as described in Supporting Text, Section 3.1. The column in $\boldsymbol{\gamma}_{\text{scan}}$ corresponding to such a clipped component is therefore deleted, and its values are added to $\boldsymbol{\delta}_{\text{scan}}$. The criterion for determining whether a component is clipped is chosen as whether its $2\sigma$ error ellipse (containing around ~95% of localizations) is completely inside the subregion or not.

The resulting $\boldsymbol{\gamma}_{\text{scan}}$ does not provide a complete description of the Gaussian mixture in the full dataset due to the deletion of components that are clipped during the scanning procedure. However, each clipped component that is deleted from a certain scan is, in theory, identified in at least one of the three other scans (see Fig. S2). The second step therefore consists of merging the $\boldsymbol{\gamma}_{\text{scan}}$ matrices of the four different scans. For this purpose, the Pearson correlation between the posterior probabilities of each pair of components $i$ and $j$ belonging to different scans is calculated:

$$\rho_{ij} = \frac{\sum_n(\gamma_{in} - \overline{\gamma_{in}})\left(\gamma_{jn} - \overline{\gamma_{jn}}\right)}{\sqrt{\sum_n(\gamma_{in} - \overline{\gamma_{in}})^2 \sum_k\left(\gamma_{jn} - \overline{\gamma_{jn}}\right)^2}} \qquad (20)$$

The sum runs over all localizations $n$ that have a non-zero posterior probability (e.g. excluding all localizations outside subregion $i$ and $j$). The correlation is tested against the null hypothesis that the posterior probabilities of components $i$ and $j$ are not correlated (i.e. it is verified that the correlation is larger than the values described by a simulated null hypothesis distribution). Two components identified in two different scans are considered to be identical if their correlation is significant according to the null hypothesis and if the correlation is larger than any other significant correlation involving either $i$ or $j$. After identifying all identical components, their posterior probabilities are combined by averaging, while the posterior probabilities of components identified in only one scan are retained. This results in a final $\boldsymbol{\gamma}$ that describes the full dataset without clipped components. The background posterior probabilities are combined similarly into a final $\boldsymbol{\delta}$.

# 4. Merging procedure

The merging procedure illustrated in Fig. 5A is performed by splitting the mixture components obtained by EMGM into two categories: the ones whose $1\sigma$ error ellipse intersects with at least one other error ellipse, called the "overlapping" components, and the ones whose $1\sigma$ error ellipse does not intersect with another one, called the "isolated" components. The $1\sigma$ error ellipse is chosen because it corresponds to the probability of containing ~40% of all localizations. This means that localizations on the intersection between two such error ellipses have approximately an equal probability to belong to both corresponding components, therefore suggesting that they can be viewed as a single merged object. Once a set of $K_{\text{overlap}}$ overlapping components have been verified, a new merged object can be calculated by summing their posterior probabilities $\gamma_{nk}$ (see Eq. (3)):

$$\gamma_{n,\text{merged}} = \sum_{k=1}^{K_{\text{overlap}}} \gamma_{nk} \tag{21}$$

The properties of the merged object can then be calculated using Eq. (4). This gives rise to a third category, called the "merged" components.

## 5. PAINT imaging of integrin β3

### 5.1 Sample preparation

We used a commercial kit (Ultivue-2, Ultivue) for our points accumulation in nanoscale topography (PAINT) [7] experiments. The sample was prepared according to the manufacturer's recommendations. Briefly, we seeded around $10^5$ REF cells on a fibronectin-coated 25 mm diameter cover slip, incubated them at 37° C in cell culture medium, washed them with PBS after 24h, and fixed them with 2.5% paraformaldehyde at 37° C for 10 minutes (see *Materials and Methods*). After removing the fixative, the cells were washed three times with PBS, the cover slip was placed into a custom made holder, and they were incubated in PBS for 10 minutes at 37° C.

The cells were subsequently reduced by incubating them for 10 minutes in a freshly prepared 0.1% sodium borohydride solution at room temperature. Afterwards, the cells were washed three times with PBS, and incubated in PBS for 10 minutes at room temperature. Next, the cells were incubated for 1.5h at room temperature in a blocking and permeabilization buffer consisting of PBS with 3% bovine serum albumin and 0.2% Triton X-100.

The primary antibody staining was carried out by incubating the cells overnight at 4 °C with integrin β3 mouse monoclonal antibodies (sc-7311, Santa Cruz Biotechnology) diluted 100 times in staining buffer composed of PBS with 1% bovine serum albumin and 0.2% Triton X-100. Next, the cells were washed four times with PBS, and incubated in PBS for 10 minutes at room temperature. The secondary antibody staining was carried out by first incubating the cells in Antibody Dilution Buffer (Ultivue-2, Ultivue) for 10 minutes at room temperature, and then for 2h with Goat-anti-Mouse-D1 antibodies (Ultivue-2, Ultivue) diluted 100 times in Antibody Dilution Buffer. Next, the cells were washed four times with PBS, and incubated in PBS for 10 minutes at room temperature.

### 5.2 Imaging procedure

Prior to imaging, 100 nm gold nanospheres (C-AU-0.100, Corpuscular) were added to the sample for lateral drift correction (see *Materials and Methods*). Imaging was performed using image strand I1-560 (Ultivue-2, Ultivue) diluted in Image Buffer (Ultivue-2, Ultivue) at a concentration of 1 nM. The imaging procedure was similar as for the PALM measurements (see *Materials and Methods*).

### 5.3 Discussion

We used PAINT to image fixed rat embryonic fibroblast (REF) cells where integrin β3 was antibody stained. The resulting PAINT images show FAs as patchy structures (Fig. S9). We hypothesize that this is caused by difficulties in labeling integrin with antibodies, for instance due to cell membrane areas that are curved inwards, resulting in an integrin epitope that is more difficult to access. We also noticed that mostly the cell periphery was labelled, again suggesting that not all integrins are accessible for the antibodies.

## 6. Production of nano-patterned substrates

Nano-patterned substrates were prepared by means of block-copolymer micelle nanolithography (BCML) as previously described [8-10]. Briefly, quasi-hexagonally ordered gold nanoparticles arrays on cleaned 25 mm diameter microscope cover slips (#1.5 Micro Coverglass, Electron Microscopy Sciences) were fabricated using toluene solution of poly(styrene)-block-poly(2-vinyl pyridine) (PS-b-P2VP, Polymer Source Inc.) [9, 10]. The PS-b-P2VP toluene solution was treated with $HAuCl_4$ (Sigma Aldrich) at a stoichiometric loading of ($P2VP/HAuCl_4$) = 0.5 and stirred at least for 24h in order to obtain gold nanoparticles (AuNPs) with a diameter between 6-8 nm. The lateral distance between the individual AuNPs was adjusted by varying the micellar coating process (spinning speed). Details concerning the applied block polymers and the spin casting processes are included in Table S1.

The area between the AuNPs was passivated with PLL-g-PEG (PLL(20kDa)-g[3.5]-PEG(2kDa), Susos AG) to prevent non-specific adhesion. The substrates were first activated in an oxygen plasma at 0.4 mbar and 150 W for 10 minutes. The PLL-g-PEG was diluted to a concentration of 0.25 mg/ml in a 10 mM HEPES buffer at pH 7.4. The freshly activated substrates were incubated upside down for 45 minutes at room temperature on a 60 µl drop of the PLL-g-PEG solution on parafilm in a moist chamber. Afterwards the substrates are washed once with milli-Q water. Following passivation, each surface was functionalized with cRGD pentapeptide (Peptide Specialty Laboratories GmbH) at a concentration of 25 µM in MilliQ water for 2h at room temperature. The cRGD pentapeptide was conjugated with a PEG spacer (6 units) that serves as a breach between the peptide and the cysteine. The physisorbed material was removed by thorough rinsing with MilliQ water and PBS.

## Supporting Figures

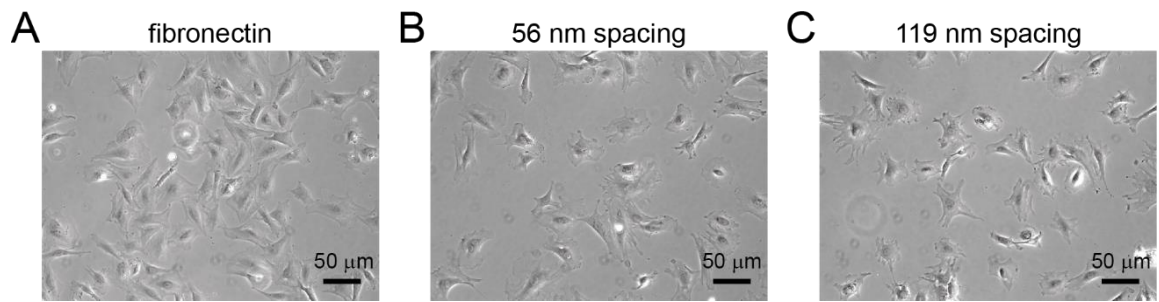

**Figure S1.** Dark field microscopy imaging of REF cells. (A-C) The REF cells were growing on (A) a fibronectin coated substrate, (b) a nano-patterned substrate with 56 nm spacing between AuNPs, or (C) a nano-patterned substrate with 119 nm spacing between AuNPs. The images were recorded 24h after transection with the integrin β3 vector.

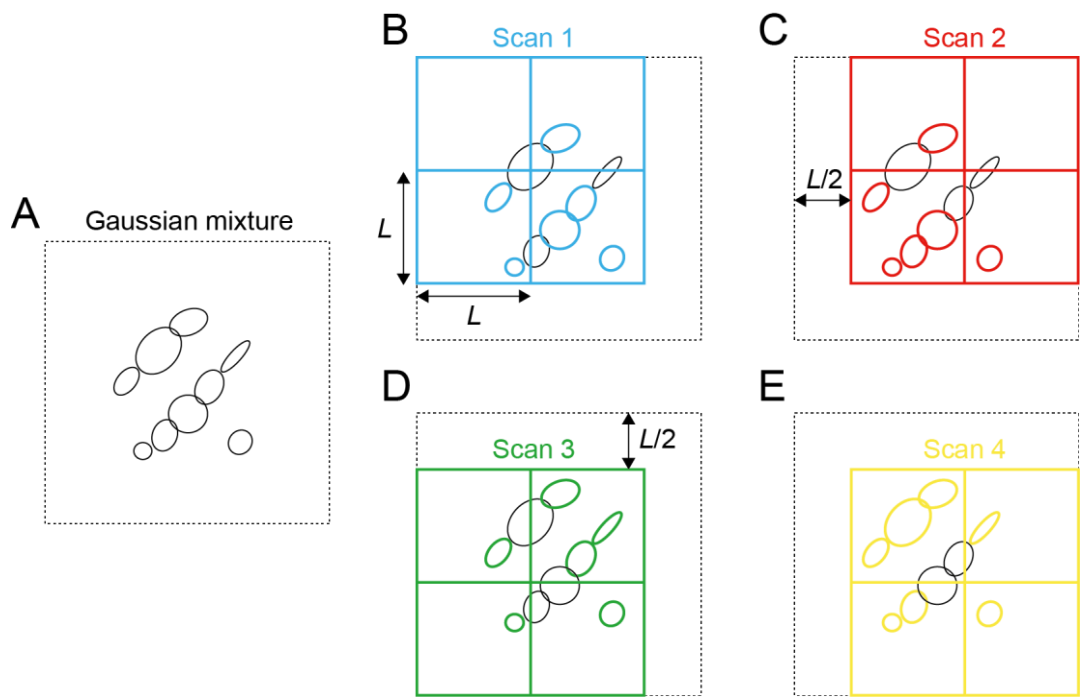

**Figure S2.** Scanning procedure for EMGM analysis of SMLM data. (A) Illustration of a Gaussian mixture with components represented by black ellipses. (B-E) Scanning procedure consisting of 4 different scans. During each scan, the EMGM analysis is performed on separate square subregions with a side length $L$, indicated by the colored squares. The Gaussian mixture components that can be correctly identified in a certain scan are indicated by the ellipses that have the same color as the squares. In between scans, the subregions are shifted over a distance $L/2$ in one of the following directions: left, right, up, or down.

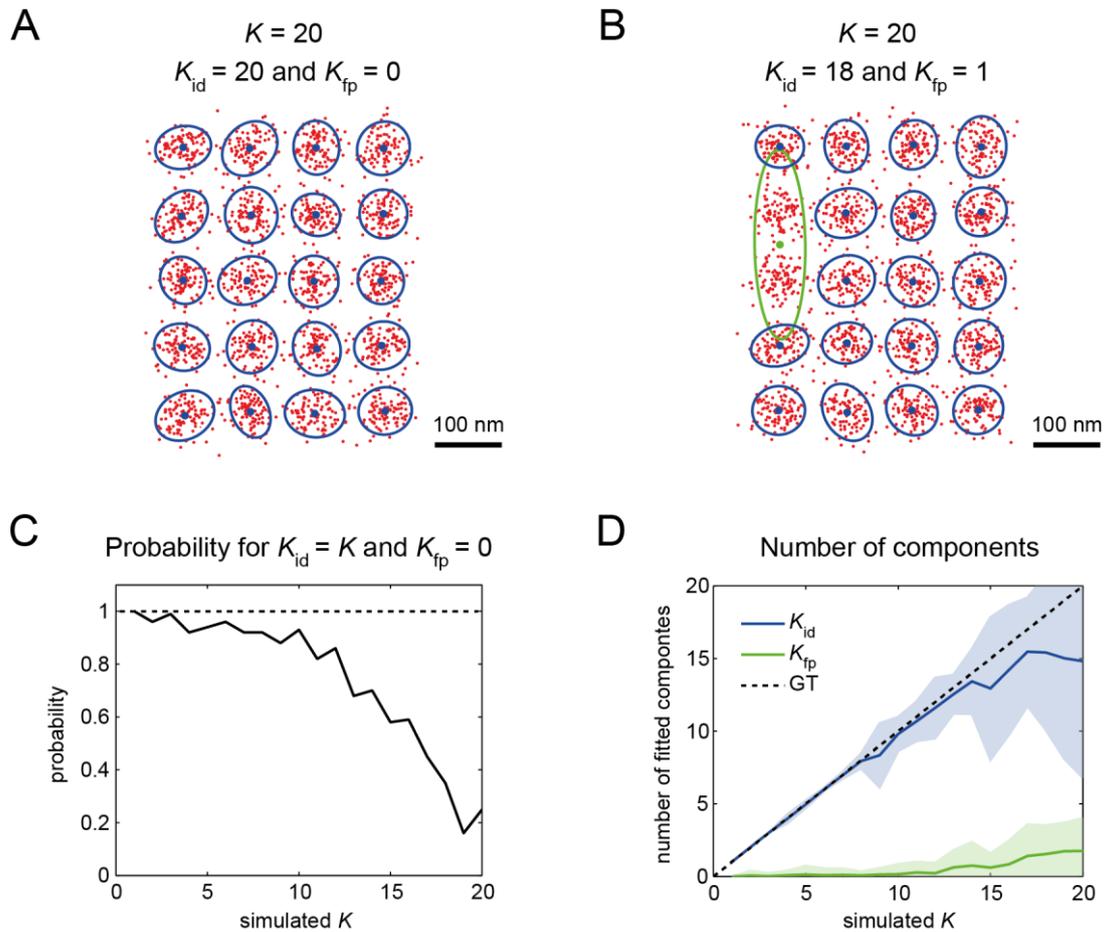

**Figure S3.** Influence of the number of mixture components $K$ on the EMGM performance. (A-B) Example EMGM results for simulated Gaussian mixtures with $K = 20$ components. EMGM correctly identified $K_{id} = 20$ components and found $K_{fp} = 0$ false positive components for (A). EMGM correctly identified $K_{id} = 18$ components and found $K_{fp} = 1$ false positive components for (B). The red dots symbolize the simulated localizations. The blue/green dots symbolize the center positions of the correct/false positive components, the blue/green ellipses symbolize the $2\sigma$ error ellipses of the correct/false positive components. (D) The simulated probability of obtaining a completely correct EMGM result (i.e. $K_{id} = K$ and $K_{fp} = 0$) as a function of $K$. (E) The simulated average values of $K_{id}$ and $K_{fp}$ as a function of $K$. The dashed line represents the ground truth (GT) and the shaded areas the standard deviation ($n = 100$).

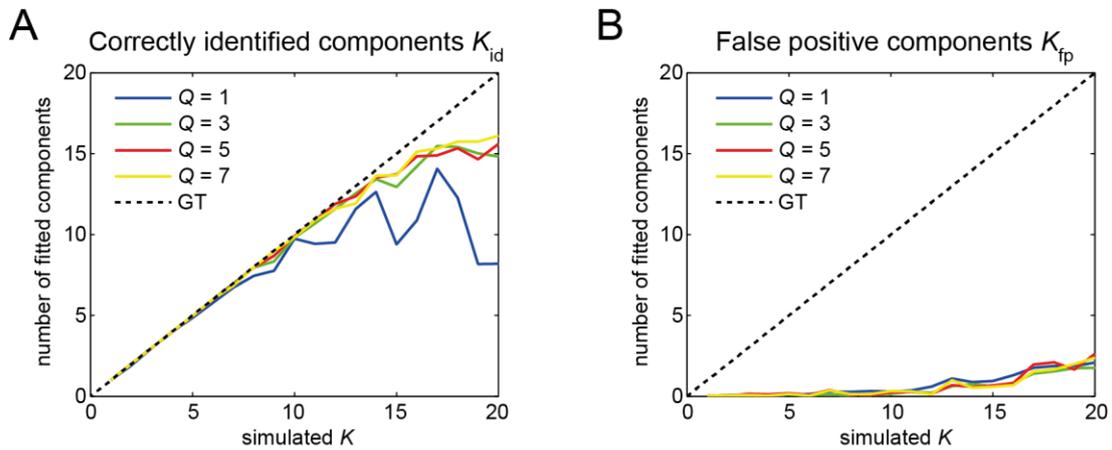

**Figure S4.** Influence of the number of initialization procedures $Q$ on the EMGM performance. Gaussian mixtures with different values of $K$ were simulated and analyzed by EMGM. (A) The average value of the number of correctly identified compenents $K_{id}$ as a function of $K$, for different values of $Q$. (B) The average value of number of false positive components $K_{fp}$ as a function of $K$, for different values of $Q$. The dashed line represents the ground truth (GT) and the shaded areas the standard deviation ($n = 100$).

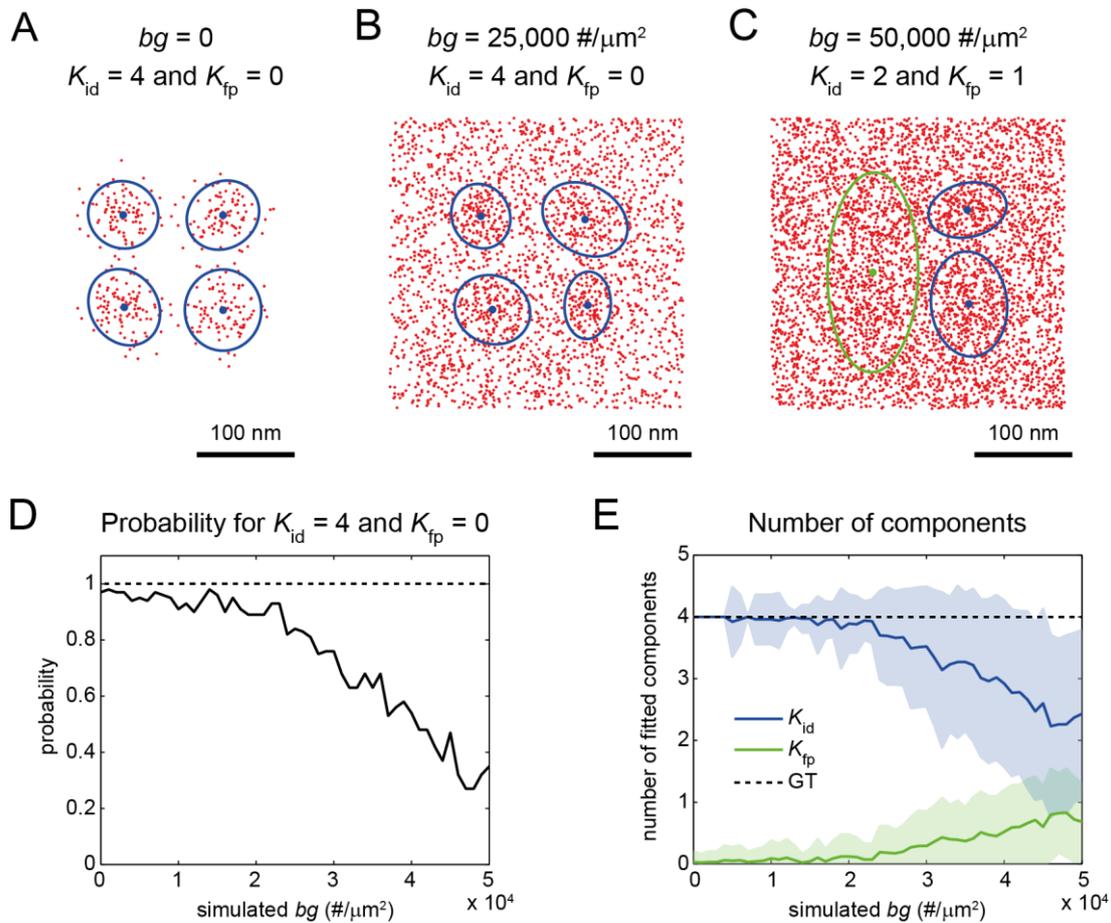

**Figure S5.** Influence of the localization background on the EMGM performance. (A-C) Example EMGM results for simulated Gaussian mixtures. Each mixture consists of $K = 4$ components with localization background density (A) $bg = 0$, (B) $bg = 25,000$ #/μm², or (C) $bg = 50,000$ #/μm². EMGM correctly identified $K_{id} = 4$ components and found $K_{fp} = 0$ false positive components for (A) and (B). EMGM correctly identified $K_{id} = 2$ components and found $K_{fp} = 1$ false positive component for (C). The red dots symbolize the simulated localizations. The blue/green dots symbolize the center positions of the correct/false positive components, the blue/green ellipses symbolize the $2\sigma$ error ellipses of the correct/false positive components. (D) The simulated probability of obtaining a completely correct EMGM result (i.e. $K_{id} = 4$ and $K_{fp} = 0$) as a function of $bg$. (E) The simulated average values of $K_{id}$ and $K_{fp}$ as a function of $bg$. The dashed line represents the ground truth (GT) and the shaded areas the standard deviation ($n = 100$).

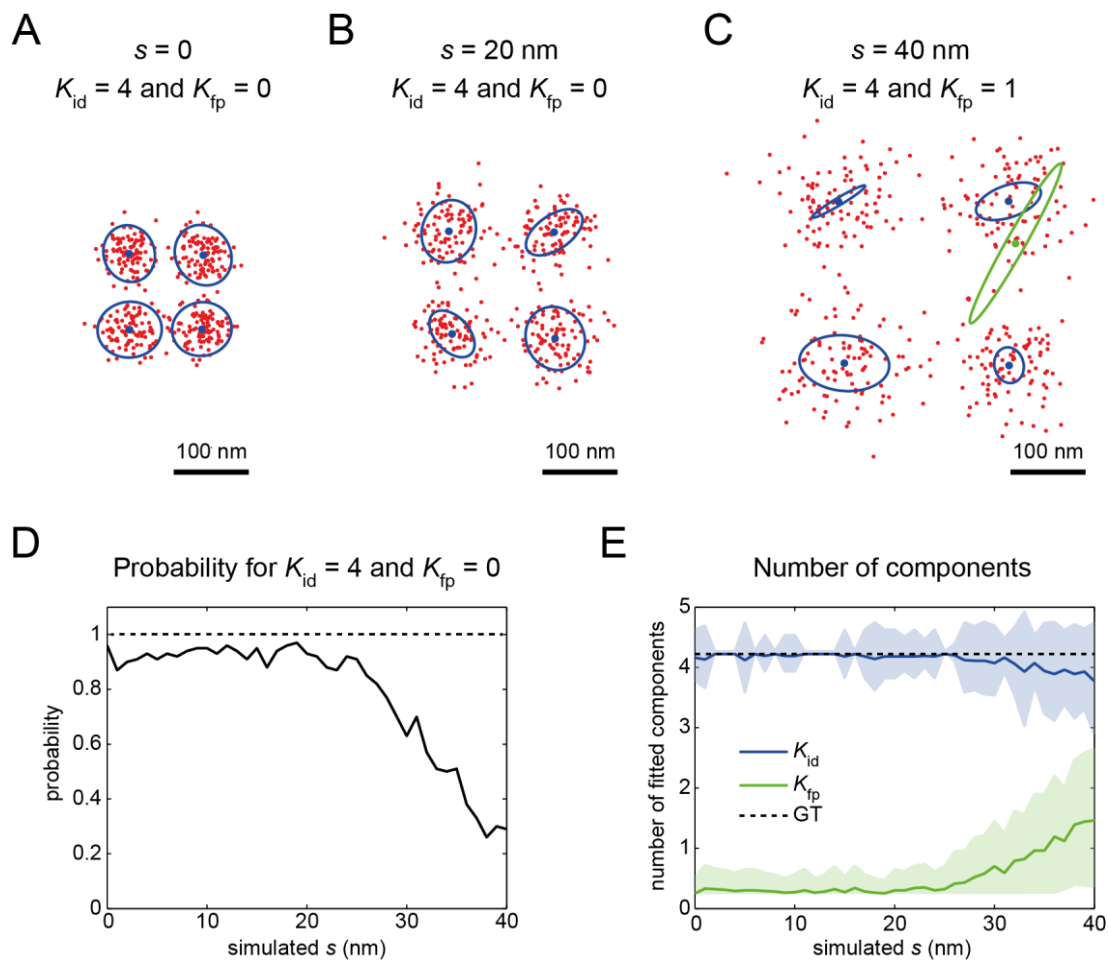

**Figure S6.** Influence of the localization uncertainty on the EMGM performance. (A-C) Example EMGM results for simulated Gaussian mixtures. Each mixture consists of $K = 4$ components with localization uncertainty (A) $s = 0$, (B) $s = 20$ nm, or (C) $s = 40$ nm. EMGM correctly identified $K_{id} = 4$ components and found $K_{fp} = 0$ false positive components for (A) and (B). EMGM correctly identified $K_{id} = 4$ components and found $K_{fp} = 1$ false positive component for (C). The red dots symbolize the simulated localizations. The blue/green dots symbolize the center positions of the correct/false positive components, the blue/green ellipses symbolize the $2\sigma$ error ellipses of the correct/false positive components. (D) The simulated probability of obtaining a completely correct EMGM result (i.e. $K_{id} = 4$ and $K_{fp} = 0$) as a function of $s$. (E) The simulated average values of $K_{id}$ and $K_{fp}$ as a function of $s$. The dashed line represents the ground truth (GT) and the shaded areas the standard deviation ($n = 100$).

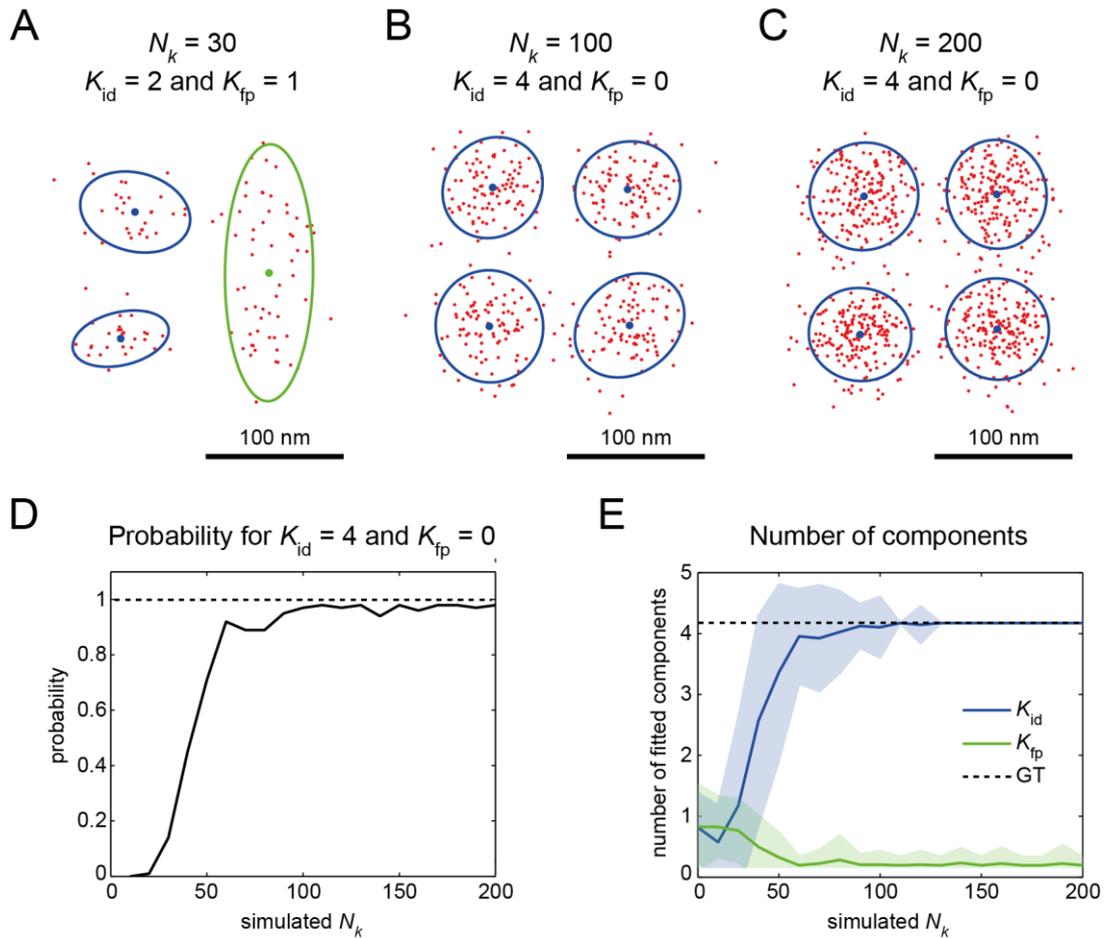

**Figure S7.** Influence of the number of localizations on the EMGM performance. (A-C) Example EMGM results for simulated Gaussian mixtures. Each mixture consists of $K = 4$ components with localization number (A) $N_k = 30$, (B) $N_k = 100$, or (C) $N_k = 200$. EMGM correctly identified $K_{id} = 2$ components and found $K_{fp} = 1$ false positive components for (A). EMGM correctly identified $K_{id} = 4$ components and found $K_{fp} = 0$ false positive component for (B) and (C). The red dots symbolize the simulated localizations. The blue/green dots symbolize the center positions of the correct/false positive components, the blue/green ellipses symbolize the $2\sigma$ error ellipses of the correct/false positive components. (D) The simulated probability of obtaining a completely correct EMGM result (i.e. $K_{id} = 4$ and $K_{fp} = 0$) as a function of $N_k$. (E) The simulated average values of $K_{id}$ and $K_{fp}$ as a function of $N_k$. The dashed line represents the ground truth (GT) and the shaded areas the standard deviation ($n = 100$).

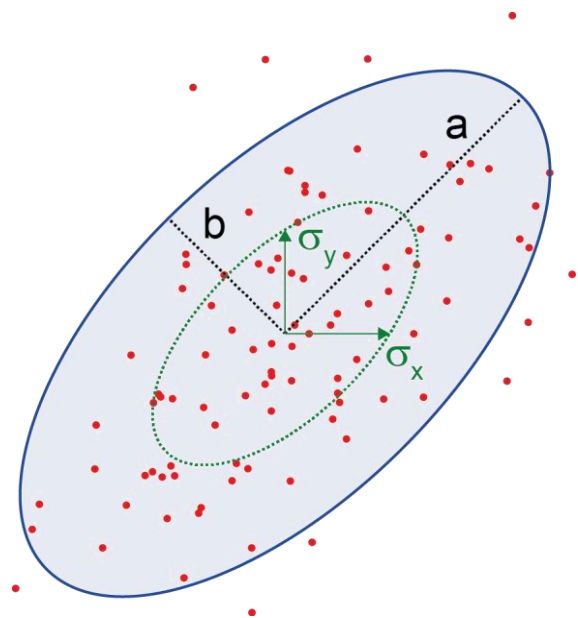

**Figure S8.** Illustration of a Gaussian component with standard deviation $\sigma_x$ and $\sigma_y$, together with the corresponding $2\sigma$ error ellipse with major axis $a$ and minor axis $b$.

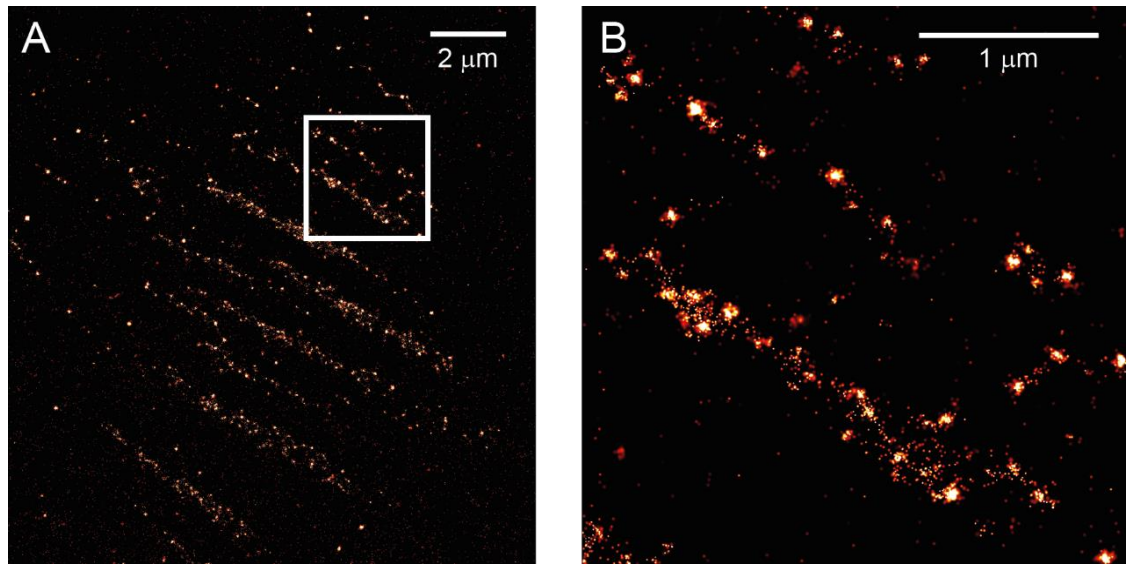

**Figure S9.** PAINT imaging of focal adhesions. (A) PAINT image of a fixed REF cell where integrin β3 was antibody stained. (B) Zoom-in of the region in (A) indicated by the white rectangle.

# Supporting Tables

| Polymer $PS_{(units)}$-b-$P2VP_{(units)}$ | PDI | Polymer concentration [mg/ml] | Spinning speed | Distance on glass [nm] |
|---|---|---|---|---|
| $PS_{1056}$-*b*-$P2VP_{671}$ | 1.09 | 5 | 2000 rpm | $56 \pm 9$ |
| | | 2.5 | 6000 rpm | $119 \pm 11$ |

**Table S1.** Details concerning the block polymers and the spin casting processes used for the fabrication of the nano-patterned substrates.